%% file: svm2.tex
\documentclass[aps,pre,pdftex,superscriptaddress,twocolumn,tightenlines,showpacs,floatfix,amsmath,amssymb,usenames,dvipsnames]{revtex4-1}

\usepackage{array}
\usepackage{amsmath}
\usepackage{amssymb}
\usepackage{bm}
\usepackage{bbm}
\usepackage{float}
\usepackage{tikz} 
\usepackage{graphicx}
\usepackage{booktabs}
\usepackage[normalem]{ulem}
\usepackage{float}
\usepackage{mathtools}
\usepackage{hyperref}
\usepackage{siunitx}
\usepackage{gnuplot-lua-tikz}
\ifcsname gpsetdashtype\endcsname
\else
\newcommand{\gpsetdashtype}[1]{} 
\fi

\usepackage[outdir=./]{epstopdf}

\usepackage[caption=false]{subfig}

\def\id{\mathbbm{1}}

\newcommand{\mean}[1]{\langle #1\rangle}
\newcommand{\mb}[1]{\mathbf{#1}}
\newcommand{\mbb}[1]{\mathbb{#1}}
\newcommand{\bds}[1]{\boldsymbol{#1}}

\newcommand{\f}[2]{\frac{#1}{#2}}

\newcommand{\innerprod}[2]{#1 \cdot #2}

\newcommand*\circled[1]{\tikz[baseline=(char.base)]{
            \node[shape=circle,draw,inner sep=1pt] (char) {#1};}}

\newcolumntype{C}{>{$}c<{$}}
\AtBeginDocument{
\heavyrulewidth=.08em
\lightrulewidth=.05em
\cmidrulewidth=.03em
\belowrulesep=.65ex
\belowbottomsep=0pt
\aboverulesep=.4ex
\abovetopsep=0pt
\cmidrulesep=\doublerulesep
\cmidrulekern=.5em
\defaultaddspace=.5em
}

\begin{document}

\title{Learning multiple order parameters with interpretable machines}
\date{\today}

\author{Ke Liu}
\email{ke.liu@lmu.de}
\affiliation{Arnold Sommerfeld Center for Theoretical Physics,
Munich Center for Quantum Science and Technology,
University of Munich, Theresienstrasse 37, 80333 M\"unchen, Germany}

\author{Jonas Greitemann}
\affiliation{Arnold Sommerfeld Center for Theoretical Physics,
Munich Center for Quantum Science and Technology,
University of Munich, Theresienstrasse 37, 80333 M\"unchen, Germany}

\author{Lode Pollet}
\affiliation{Arnold Sommerfeld Center for Theoretical Physics,
Munich Center for Quantum Science and Technology,
University of Munich, Theresienstrasse 37, 80333 M\"unchen, Germany}

\begin{abstract}
Machine-learning techniques are evolving into a subsidiary tool for studying phase transitions in many-body systems.
However, most studies are tied to situations involving only one phase transition and one order parameter.
Systems that accommodate multiple phases of coexisting and competing orders, which are common in condensed matter physics, remain largely unexplored from a machine-learning perspective.
In this paper, we investigate multiclassification of phases using support vector machines (SVMs) and apply a recently introduced kernel method for detecting hidden spin and orbital orders to learn multiple phases and their analytical order parameters.
Our focus is on multipolar orders and their tensorial order parameters whose identification is difficult with traditional methods.
The importance of interpretability is emphasized for physical applications of multiclassification.
Furthermore, we discuss an intrinsic parameter of SVM, the bias, which allows for a special interpretation in the classification of phases, and its utility in diagnosing the existence of phase transitions.
We show that it can be exploited as an efficient way to explore the topology of unknown phase diagrams where the supervision is entirely delegated to the machine.
\end{abstract}

\maketitle

\section{Introduction}
Order parameters play a central role in condensed matter physics~\cite{BookAnderson, Mermin79}.
They represent the important degrees of freedom in a system, without tracking all the microscopic details.
They define states of matter and identify their elementary excitations and topological defects.
Recognizing the order parameter is often a basic step toward understanding the physics and constructing the effective theory of a many-body system.

For some systems, the order parameter is obvious. For example, in ferromagnets, the order parameter, namely the magnetization, is simply given by a sum of local magnetic moments.
In general, however, identifying the order parameter can be a difficult task.

This is particularly the case in frustrated spin and orbital systems~\cite{BookLacroix}.
For instance, the classical Heisenberg antiferromagnet on a kagome lattice was thought to be a candidate for a classical spin liquid.
However, it was later realized that its low-temperature phase is actually a spin nematic:
owing to an order-by-disorder phenomenon, it possesses a hidden quadrupolar order~\cite{Chalker92, Reimers93} as well as an (in-plane) octupolar order~\cite{Ritchey93, Zhitomirsky02, Zhitomirsky08}.
The subtlety of identifying these orders may be partially understood in the following way:
they appear as collective motion and higher moments of ordinary spins.
To define their order parameter, one needs to recognize the relevant degrees of freedom and the pattern in which those are combined.

The quadrupolar and octupolar orders in the above kagome antiferromagnet are instances of multipolar orders.
Such orders also commonly occur in other frustrated systems, such as pyrochlores~\cite{Yamaura12, Takatsu16, TaillefumierShannon17,YanShannon17}, garnets~\cite{Paddison15}, triangular antiferromagnets~\cite{MomoiShannon06}, and frustrated orbital systems~\cite{Orlova17, Wu08, ChernWu14}.
However, no general guiding principle is known to predict their existence and to identify them.

Recent developments in physical applications of machine-learning techniques may provide a means to detect order parameters of many-body systems.
The most popular algorithms for this include a variety of neural networks~\cite{CarrasquillaMelko17, Nieuwenburg17, BroeckerTrebst17, Khatami17, LiuYH18, Beach18, Suchsland18} and kernel methods~\cite{Wang16, WangZhai17, PonteMelko17}.
Algorithms of the former type, such as restricted Boltzmann machines (RBMs), convolutional neural networks (CNNs), and generative adversarial networks (GANs), feature a high expressiveness.
Given a sufficient number of hidden units, they can in principle represent any nonlinear function.
However, these methods face problems in their interpretability: except for the simplest models, it is in general very difficult to understand the machine results.
On the other hand, the kernel methods, such as support vector machines (SVMs) and kernel principal component analysis (kernel PCA), are strongly interpretable, which permits theoretical insight from machine results.
Nevertheless, the scope of their applicability crucially relies on the choice of the kernel function, and standard SVM and PCA kernels may only apply to simple physical problems.

The trade-off between expressiveness and interpretability is a prevalent theme in machine learning.
For detecting hidden multipolar orders in frustrated systems, it is possible to reconcile them.
In Ref.~\cite{Greitemann19}, we have introduced a kernel for SVMs that can be used to probe general classical $\mathrm O(3)$-breaking orientational order.
We demonstrated its capabilities by learning the analytical order parameter of multipolar orders up to rank $6$ and discussed its application to the identification of novel spin nematics and for ruling out spurious spin-liquid candidates.

In Ref.~\cite{Greitemann19}, as in most studies related to the machine learning of phases, the focus is on situations where only two phases (defining one order parameter) are involved, and a binary classification is achieved.
In this paper, we apply our kernel to multiclassification to handle multiple phases. Moreover, we also introduce a criterion which can be used to explore the initially unknown phase diagram.

Although the classification of multiple labels is a standard extension to SVMs, the physical application is far from trivial and demands strong interpretability.
This can be explained in the following way:
one obtains numerical classifiers that each distinguish a pair of labels (representing two phases).
If a system accommodates multiple phases, we need to interpret those classifiers to discern and define each phase in physical language.
Moreover, if a phase possesses a number of orders simultaneously, the machine will combine them into a single classifier. In order to correctly understand the underlying physics, we need to be able to isolate them individually.

The remainder of the paper is organized as follows.
In Sec.~\ref{sec:SVM}, we review some basic concepts of SVM, including a brief introduction of binary and multiclassification.
In Sec.~\ref{sec:multipolar}, we review the construction of our kernel, and discuss the bias parameter in SVM and its expected value as applied to phase classification.
In Sec.~\ref{sec:gauge}, we define a model that generates multipolar-ordered spin configurations serving as training and testing data.
In Sec.~\ref{sec:Dinfh}, we take a simple quadrupolar order as example, and explain in detail the procedure employed to extract the analytical order parameter and the use of the aforementioned bias parameter.
Moreover, in this section we also compare different choices of training data. This discussion may also be applicable to other machine-learning methods.
In Sec.~\ref{sec:biaxial}, we use our method to detect multiple (multipolar) phases and learn their order parameters.
A generalized quadrupolar and the octupolar order will be considered as examples.
In Sec.~\ref{sec:mapping_PD}, we examine the ability of SVM to explore the phase diagram in the absence of prior knowledge on its topology.
Our motivation there is to obtain a tentative phase diagram, and use it to guide the subsequent learning of the order parameters.
The precise phase diagram can be obtained after the order parameters have been extracted.
However, as will be discussed, the scheme is actually capable of achieving a decent precision before the order parameters are even considered.
We conclude with an outlook in Sec.~\ref{sec:conclusion}.

\section{Support Vector Machines} \label{sec:SVM}
In this section we will provide a brief introduction of SVM.
A comprehensive discussion can be found in Ref.~\cite{BookVapnik}.
The key message here is that whether SVM is able to distinguish two classes depends on the choice of the kernel function.
Readers who are already familiar with SVM or are not interested in the technical details may skip to Sec.~\ref{sec:multipolar} where we construct the kernel suited for the physical problem.

\subsection{Definition of the optimization problem}

Support vector machines provide a means to construct a classifier from
$N_s$ training data samples, $(\mb x^{(k)}, y^{(k)})$, where $\mb
x^{(k)}\in\mathbb{R}^d$ and $y^{(k)}$ labels the class of the training sample.
We will consider the case of binary classification, $y^{(k)}\in\{-1,+1\}$, first and explain the extension to more than two classes afterwards~\cite{CortesVapnik95,Chang01,Chang11}.

The aim of the SVM is to find a hyperplane, dubbed decision boundary,
\begin{align}
  \innerprod{\mb{w}}{\mb{x}} - \rho = 0,
\end{align}
defined by parameters $\mb w\in\mathbb{R}^d$ and $\rho\in\mathbb{R}$, that
separates the data into the two classes. If the data are separable, typically infinitely many
solutions exist. In order to select the solution that separates the data most
clearly, a finite margin is imposed which must not contain any of the training
data samples, and the width of the margin is sought to be maximal. When defining
$\innerprod{\mb{w}}{\mb{x}} - \rho = \pm 1$ as the boundaries of the margin,
such that the width of the margin is given by $2/\|w\|$, the optimization
problem may then be written as
\begin{align}
  \begin{cases}
    \text{minimize} & \frac{1}{2}\|\mb w\|^2\quad\text{w.r.t.}\quad\mb{w},\rho,\\
    \text{subject to} & y^{(k)}\left( \innerprod{\mb{w}}{\mb{x}^{(k)}} - \rho \right) \ge 1 \quad\forall\ k.
  \end{cases}
  \label{eq:hard_margin}
\end{align}

Most of the time, the data may not be perfectly separable and the constraint
\eqref{eq:hard_margin} must be relaxed, allowing for incursions into the
margin at a cost proportional to $C>0$:
\begin{align}
  \begin{cases}
    \text{min.} & \frac{1}{2}\|\mb w\|^2+C\sum_k\xi_k\quad\text{w.r.t.}\quad\mb{w},\rho,\{\xi_k\},\\
    \text{subj.} & y^{(k)}\left( \innerprod{\mb{w}}{\mb{x}^{(k)}} - \rho \right) \ge 1-\xi_k,\ \ \xi_k\ge 0 \ \ \forall\ k
  \end{cases}
  \label{eq:soft_margin}
\end{align}
where $\xi_k$ are slack variables. $\xi_k> 1$ corresponds to a misclassified
sample.
Since the training data enter the optimization problem only as inner
products in the constraint, the optimal $\mb w$ will lie in the span of the
training data,
\begin{align}
  \mb w = \sum_k \lambda_ky^{(k)}\mb x^{(k)}.\label{eq:SVM_w}
\end{align}
Indeed, only those samples which violate (or touch) the margin necessitate a nonzero slack variable and contribute to $\mb w$ with a nonzero $\lambda_k$.
These are the support vectors.

Larger values of $C$ ``harden'' the margin, making it more narrow and reducing the
number of support vectors. This comes at the risk of overfitting noise. In the
limit $C\to\infty$, the solution to Eq.~\eqref{eq:hard_margin} is reproduced.
Smaller values of $C$ regularize the problem more strongly,
allowing for a wider margin and consequently a higher number of support vectors,
but potentially discard faithful information in the training data set.

The optimal choice of the regularization parameter $C$ is ultimately
problem specific. In principle, values of $C$ can span many
orders of magnitude.
There exists, however, an alternative reparametrization of
Eq.~\eqref{eq:soft_margin} in terms of a regularization parameter $\nu\in[0,1)$
which has been shown to impose an upper bound on the fraction of training
samples that violate the margin and a lower bound on the number of training
samples that serve as support vectors \cite{Scholkopf00}. $\nu$-SVM thus
admits a more universal interpretation and we found it to simplify the selection
of an appropriate regularization. The optimization problem may not have a
feasible solution for all values of $\nu$ if the training data are unbalanced.
In fact, the maximum feasible value is given by $\nu_\text{max}=2\min(N_s^+,
N_s^-)/N_s$ where $N_s^\pm$ are the number of training samples in either class \cite{Chang01}.


\subsection{Decision function}

The quadratic programming problem \eqref{eq:soft_margin} can be numerically
solved using standard methods such as sequential minimal optimization~\cite{Platt98,Fan05}.
The complexity of this algorithm is somewhat dependent on the nature of the training data. Reference~\cite{Platt98} gives empirical data of better than $\mathcal{O}(N_s^{2.2})$ which is roughly in line with our observations. Note that in this work, the computational effort was dominated by the generation of independent samples, rather than the SVM optimization, in all cases but Sec.~\ref{sec:mapping_PD}.

The decision function
\begin{align}
  d(\mb x) = \innerprod{\mb w}{\mb x} - \rho\label{eq:decfun}
\end{align}
determines the orientated distance of a test sample $\mb{x}$ from the hyperplane and its sign can be used to predict the class label. Plugging
Eq.~\eqref{eq:SVM_w} into the above allows for the practical calculation of the
decision function as a sum over inner products with the support vectors:
\begin{align}
  d(\mb x) = \sum_k\lambda_ky^{(k)}\innerprod{\mb x^{(k)}}{\mb x}-\rho.\label{eq:decfun_supp}
\end{align}

The solution to the optimization problem is thus entirely given by the optimal
variables $\{\lambda_k\}$ and the bias $\rho$, whereas the optimal values of the
slack variables are given by the so-called hinge loss,
\begin{align}
  \xi_k = \max\{0, -y^{(k)}d(\mb{x}^{(k)})\}.\label{eq:hinge}
\end{align}

\subsection{The kernel trick}\label{sec:kernel_trick}

Often, the data in the raw $d$-dimensional feature space is not expected to be
separable by a hyperplane. The solution is to invoke a mapping $\mb x\mapsto
\bds{\varphi}(\mb x)$ to a higher-dimensional auxiliary space where the mapped training
data become linearly separable. Further, since the decision function and the
optimization problem \eqref{eq:soft_margin} involves only inner products in
the auxiliary space, the details of the mapping $\bds{\varphi}$ (and even the
dimensionality of the auxiliary space) need not be known as long as one can
compute a kernel function $K$ of the original data in the raw feature
space, such that
\begin{align}
  K(\mb x, \mb y) = \innerprod{\bds{\varphi}(\mb x)}{\bds{\varphi}(\mb y)}.
\end{align}

The decision function is consequently obtained by replacing the inner product in Eq.~\eqref{eq:decfun_supp} with an evaluation of the kernel function,
\begin{align}
  d(\mb x) = \sum_k\lambda_ky^{(k)}K(\mb x^{(k)}, \mb x)-\rho.\label{eq:decfun_kernel}
\end{align}
This is also done in the definition of the optimization problem \eqref{eq:soft_margin}, where $\mb{w}$ is first replaced by its expansion in terms of support vectors, Eq.~\eqref{eq:SVM_w}. For example, in the optimization objective, one replaces
\begin{align}
  \|\mb{w}\|^2 = \mb{w}\cdot\mb{w} =&\ \sum_{k,k'}\lambda_k\lambda_{k'}y^{(k)}y^{(k')}\mb{x}^{(k)}\cdot\mb{x}^{(k')}\notag\\
  \rightarrow&\ \sum_{k,k'}\lambda_k\lambda_{k'}y^{(k)}y^{(k')}K(\mb{x}^{(k)},\mb{x}^{(k')}).
  \label{eq:kernelized_objective}
\end{align}

\subsection{Multiclassification}\label{sec:multi}

The extension of SVM to the case where $M>2$ distinct labels are assigned to the training samples is most effectively accomplished by considering all $M(M-1)/2$ pairs of labels one by one~\cite{HsuLin02}, considering only the training samples belonging to either label and solving the binary classification problems individually. This produces $M(M-1)/2$ distinct decision functions. Any training sample that contributes to any decision function by a nonzero $\lambda_k$ is considered a support vector of the multiclassification problem. Note that depending on the nature of the problem, the share of support vectors contributing to multiple decision functions may be significant, allowing for a space-efficient representation.

When it comes to predicting the label for a test sample, each decision function establishes a precedence of one label over the other. Ideally, these relations are collectively compatible with transitivity and one can unambiguously assign a label. When this is not the case, the approach to reconcile the relations followed by most SVM packages is to ``poll'' by giving one vote to each decision function and picking the label that accumulates the majority of votes~\cite{Chang11}. This approach is not well suited to situations where the labeling of the training data does not necessarily correspond to the ``physical'' reality and multiple labels in fact represent the same class, resulting in the vote being split among them. It also fails to recognize situations where the decision function is incapable of distinguishing between two labels but rather overfits noise.

For the purpose of this work, we therefore do not follow a blanket polling scheme but consider the decision functions individually and may discard them based on physical insight.

\section{Kernel for general multipolar orders}\label{sec:multipolar}

The problem of distinguishing an ordered phase from a disordered one may be viewed as a binary classification problem. The training data consist of microscopic configurations which are labeled as being from the ordered or disordered phase.

The decision function, which quantifies the distance of a sample from the decision boundary, i.e., the phase transition, serves a similar role to the magnitude of an order parameter. Given a suitable choice of the kernel, the decision function will hence reproduce the true order parameter. This relation was first pointed out by Ponte and Melko in Ref.~\cite{PonteMelko17}.
There, the authors studied the Ising model and several of its variants using a standard quadratic kernel, which is highly interpretable but only applies to linear orders such as the Ising or $XY$ magnetization.

In Ref.~\cite{Greitemann19}, we have introduced an interpretable kernel and shown it to be capable of capturing general $\mathrm O(N)$-breaking orientational orders, with $N \leq 3$.
Below we will first review the construction of this kernel and then discuss its further potential in the detection of phase transitions.

\subsection{Definition of the kernel}\label{sec:kernel}

Without assuming the specific form of a potential multipolar order, there are two basic properties we can exploit to construct our kernel:
(i) local order can be defined by a finite number of local fields; (ii) a multipolar order can generally be formulated in terms of tensors (or polynomials) with a finite rank (degree) that are invariant under certain point-group transformations~\cite{Michel01,Nissinen16}.

Hence, we can partition the system into clusters, each containing a finite number of spins,
\begin{align} \label{eq:clustering}
   \mb{x} &= \{ \mb{S}_i \} = \{ \mb{S}_{I}^\alpha \} = \{ S_{I,a}^\alpha \},
\end{align}
where $\mb{x}$ is the configuration vector and $\mb{S}_i$ are $\mathrm O(3)$ spins at lattice site $i$.
The index $I$ enumerates spin clusters and $\alpha=\rm{l, m, \dots}$ identifies spins within a cluster, such that $i=(I,\alpha)$. $a=x,y,z$ runs over the components of each spin.
We then map the spin components within each cluster to all monomials of degree $n$, and consequently perform a lattice average over all clusters,
 \begin{align} \label{eq:mapping}
 	\mb{x} \ \mapsto \ \bds{\phi}(\mb{x}) &=\{\phi_{\mu}\}= \{ \langle S^{\alpha_1}_{a_1} \dots S^{\alpha_n}_{a_n} \rangle_{cl} \},
 \end{align}
where $\mean{\dots}_{cl}$ denotes the lattice average over spin clusters $I$. We also introduce a multi-index to collectively refer to the individual monomials, $\mu=(\alpha_1,\dots,\alpha_n;a_1,\dots,a_n)$.

The kernel is then defined as
\begin{align} \label{eq:kernel}
	K\big(\mb{x}, \mb{x}^{\prime}\big) =  \big[\bds{\phi}(\mb{x}) \cdot \bds{\phi}(\mb{x}^{\prime})\big]^2.
\end{align}
Formally, it is a quadratic kernel with respect to the feature vector $\bds{\phi}(\mb{x})$ after the monomial mapping, Eq.~\eqref{eq:mapping}.
With this kernel, the decision function can be expressed as
\begin{align}
	 d(\mb{x}) &= \sum_{k} \lambda_k y_k \big[\bds{\phi} (\mb{x}^{(k)}) \cdot \bds{\phi}(\mb{x})\big]^2 - \rho \nonumber \\
	&= \sum_{\mu \nu} C_{\mu \nu} \phi_{\mu} \phi_{\nu} - \rho, \label{eq:decfun_phi}\\
	 C_{\mu \nu} &= \sum_{k} \lambda_k y_k \langle
	S^{\alpha_1}_{a_1} \dots S^{\alpha_n}_{a_n}\rangle_{cl}^{(k)} \langle S^{\alpha^{\prime}_1}_{a^{\prime}_1} \dots S^{\alpha^{\prime}_n}_{a^{\prime}_n} \rangle_{cl}^{(k)},\label{eq:coeff_matrix}
\end{align}
where $C_{\mu\nu}$ is a coefficient matrix that is calculated from the learned support vectors.

The rank-$n$ order parameter tensor, $\mathbb{O}$, can be written as a linear combination of basis tensors,
\begin{align} \label{eq:O_tensor}
  \mathbb{O} = \sum_{\bds{\alpha}} c_{\bds{\alpha}} \mb{S}^{\alpha_1} \otimes \mb{S}^{\alpha_2} \otimes. .. \otimes \mb{S}^{\alpha_n},
\end{align}
where the coordinates $c_{\bds{\alpha}}$ encode the sought after analytical structure of $\mathbb{O}$. The corresponding magnitude of the ordering is obtained by taking a tensor-analog of the Frobenius norm $\|\mathbb{O}\|_F^2=\sum_{a_1\dots a_n}|\mathbb{O}_{a_1\dots a_n}|^2$. Thus, when regarding the decision function as the magnitude of the order, we aim to write Eq.~\eqref{eq:decfun_phi} as a tensor norm square $\|\mathbb{O}\|_F^2$ and extract the definition of $\mathbb{O}$ in terms of its basis tensors, Eq.~\eqref{eq:O_tensor}. An example of this is demonstrated in Sec.~\ref{sec:Dinfh_op}.

In the above construction, we made the ansatz that the spins within the cluster are sufficient to define the underlying local order.
This is done mostly to reduce the complexity of both the optimization itself and the subsequent analysis by eliminating the scaling with system size.
Note that we do not assume the cluster to be the optimal choice to accommodate the given order.
The choice of the cluster is guided by information of the lattice geometry or the Hamiltonian.
For example, one may use a number of lattice cells as a tentative cluster.
If the cluster is chosen to be larger than necessary, one will find a reducible form of the order parameter and can likely infer the optimal cluster size.

\subsection{The bias parameter in phase classification}\label{sec:bias}

We will now investigate the role of the bias parameter $\rho$ in the decision function.
Readers who are not interested in the technical details may revisit this section later, after seeing the examples in Sec.~\ref{sec:Dinfh}.
However, the basic idea can be intuitively summarized as follows.
For a fully disordered spin configuration $\tilde{\mb{x}}$, the magnitude of the ordering and thereby the first term in the decision function Eq.~\eqref{eq:decfun_phi} vanishes, leading to $d(\tilde{\mb{x}}) = -\rho$.
Consequently, $\rho = -d(\tilde{\mb{x}}) = 1$, as (ideally) all disordered configurations $\tilde{\mb{x}}$ will fall onto the lower margin boundary.
Therefore, we may use the behavior of the bias $\rho$ as an indicator to signify the presence or absence of a phase transition:
$\rho = 1$, if the data of either label correspond to the disordered and ordered phase, respectively; $\rho \neq 1$ (with significant violations) if the samples are in fact collected from the same phase or a significant portion of samples is mislabeled.

To support the above proposition, we begin by noting that the first term in SVM's optimization objective, Eq.~\eqref{eq:soft_margin}, after being kernelized [Eq.~\eqref{eq:kernelized_objective}] with the multipolar kernel, Eq.~\eqref{eq:kernel}, amounts to the Frobenius norm of the coefficient matrix, Eq.~\eqref{eq:coeff_matrix}:
\begin{align}
  \|\mb{w}\|^2 \rightarrow&\ \sum_{k,k'}\lambda_k\lambda_{k'}y^{(k)}y^{(k')}\biggl( \sum_\mu \phi^{(k)}_\mu \phi^{(k')}_\mu \biggr)^2\notag\\
  =&\ \sum_{\mu\nu}\sum_k\lambda_ky^{(k)}\phi^{(k)}_\mu\phi^{(k)}_\nu\sum_{k'}\lambda_{k'}y^{(k')}\phi^{(k')}_\mu\phi^{(k')}_\nu\notag\\
  =&\ \sum_{\mu\nu}C_{\mu\nu}^2 = \|C\|_F^2.
\end{align}

As for the second part of the optimization objective, the optimal slack variables will assume a value given by the hinge loss, Eq.~\eqref{eq:hinge},
i.e., they satisfy their constraint by equality, or they are unnecessary and will not incur any penalty to the objective.

The data $\phi_\mu$ may obey some internal constraints (such as the normalization of spin vectors, orthogonality among spins, etc.) which allow for freedom in the choice of $C_{\mu\nu}$ and $\rho$ while keeping the decision function invariant. In particular, given some matrix $D_{\mu\nu}$ which produces merely a constant when contracted with any valid feature vector $\bds{\phi}$,
\begin{align}
  \sum_{\mu\nu}D_{\mu\nu}\phi_\mu\phi_\nu &= D_0,\label{eq:D_requirement}
\end{align}
one can transform $C_{\mu\nu}\mapsto C_{\mu\nu}+\epsilon D_{\mu\nu}$ and absorb the additional constant by $\rho\mapsto\rho - \epsilon D_0$ without affecting the values of the decision function.
Since only the decision function enters the hinge loss and the inequality constraints, SVM will choose the parameter $\epsilon$ freely in a way that minimizes the Frobenius norm $\|C+\epsilon D\|^2$. The solution will thus obey
\begin{align}
  \frac{\mathrm d}{\mathrm d\epsilon}\!\sum_{\mu\nu}(C_{\mu\nu}+\epsilon D_{\mu\nu})^2 = 2\sum_{\mu\nu}(C_{\mu\nu}+\epsilon D_{\mu\nu})D_{\mu\nu} = 0.
\end{align}
The coefficient matrix $C_{\mu\nu}$ extracted from SVM already manifests the optimal choice with respect to $D_{\mu\nu}$, thus, $\|C+\epsilon D\|$ is minimal for $\epsilon=0$ which implies
\begin{align}
  \sum_{\mu\nu}C_{\mu\nu}D_{\mu\nu} = 0.\label{eq:stationary_D}
\end{align}

One particular choice is given by $\tilde D_{\mu\nu}\coloneqq \mean{\tilde\phi_\mu\tilde\phi_\nu}_\textup{diso}$ which denotes an ensemble average over configurations $\tilde{\bds{\phi}}$ from the disordered phase. The requirement of Eq.~\eqref{eq:D_requirement} is in fact fulfilled:
\begin{align}
  \sum_{\mu\nu}\tilde D_{\mu\nu}\phi_\mu\phi_\nu = \mean{(\tilde{\bds\phi}\cdot\bds\phi)^2}_\textup{diso} = \textup{const.},
\end{align}
as the disorder average amounts to an isotropic integral over the disordered spins $\tilde{\mb{S}}$ independently and thus the fixed, arbitrary feature vector $\bds{\phi}$ can be eliminated from the integrand by a change of variables.

We now calculate the value of the decision function as it is measured in the disordered phase, i.e., we subject its argument to the same disorder average,
\begin{align}
  \mean{d(\tilde{\mb{x}})}_\textup{diso} &= \Bigl\langle \sum_{\mu\nu} C_{\mu\nu}\tilde{\phi}_\mu\tilde{\phi}_\nu - \rho \Bigr\rangle_\textup{diso}\notag\\
  &= \sum_{\mu\nu}C_{\mu\nu} \mean{\tilde\phi_\mu\tilde\phi_\nu}_\textup{diso} - \rho\notag\\
  &= \sum_{\mu\nu}C_{\mu\nu}\tilde D_{\mu\nu} - \rho = -\rho,
\end{align}
by virtue of Eq.~\eqref{eq:stationary_D}. This implies that the decision function assumes a constant value throughout the disordered phase. Indeed, local order parameters are typically zero throughout the disordered phase and pick up finite values as the transition to the ordered phase takes place. One should therefore shift the decision function by $\rho$ to interpret it as an order parameter.

Since deep in the disordered phase, the individual spins are independent, the lattice average in the definition of the feature vector $\tilde{\bds{\phi}}$ already averages over many disordered spins. 
Thus, given a sufficiently large system, the statement can be refined to $d(\tilde{\mb{x}})=-\rho$ for spin configurations $\tilde{\mb{x}}$ in the disordered phase.

The phase classification problem is distinct from generic classification problems in the sense that all the data from one class, the disordered phase, (on average) trace out an isosurface of the decision function.
The decision boundary as well as the ``upper'' and ``lower'' margin boundaries are isosurfaces of the decision function too, corresponding to values, of $0$, $+1$, and $-1$ (cf. Sec.~\ref{sec:SVM}).
Thus, the ``lower'' margin boundary will fall onto the disordered samples, i.e., $d(\tilde{\mb{x}})\approx -1$, which implies $\rho=1$.

In Sec.~\ref{sec:Dinfh}, we will illustrate the above interpretation in various scenarios.
Further, in Sec.~\ref{sec:mapping_PD}, we base our analysis of the topology of an unknown phase diagram solely on the bias.

\section{Model and samples}\label{sec:gauge}

To validate the capability of the kernel Eq.~\eqref{eq:kernel} to detect general multipolar orders, we employ the following classical Hamiltonian to generate samples~\cite{Liu16}:
\begin{align} \label{eq:gauge_model}
	H = \sum_{\langle i,j \rangle} J^{\alpha \beta}_{ab}S^{\alpha}_{i,a} U^{\beta \gamma}_{ij} S^{\gamma}_{j,b}.
\end{align}
Here, $\langle i,j\rangle$ enumerates nearest-neighboring sites on a cubic lattice. $S_{i,a}^\alpha$ denotes the components $a\in\{x,y,z\}$ of three orthogonal $\mathrm O(3)$ spins at lattice site $i$, which are distinguished by a color index $\alpha \in \{\rm{l,m,n}\}$.
These spins form a local triad that can represent general spin orientations. In the notation of Sec.~\ref{sec:kernel}, these triads constitute the spin clusters.
$J^{\alpha\beta}_{ab}$ denotes a general exchange interaction.
Einstein summation over color and component indices is implied.

The Hamiltonian~\eqref{eq:gauge_model} is reminiscent of the general exchange interaction of nearest-neighboring spins,
$H_{\rm ex} = \sum_{\langle i,j \rangle} J_{ab} S_{i,a} S_{j,b}$.
However, in Eq.~\eqref{eq:gauge_model}, there are additional fields $U_{ij}$ living on the bonds $\langle i,j \rangle$. These are rotation matrices realizing elements of a three-dimensional point group $G$, $U_{ij} \in G \subset \mathrm O(3)$, and rotate the color indices of a spin, $S^{\alpha}_a = U^{\alpha \beta} S^{\beta}_a$.
These fields are in fact gauge fields, and the Hamiltonian~\eqref{eq:gauge_model} possesses a local point-group symmetry,
\begin{align}
  S^{\alpha}_{i,a} &\mapsto \Lambda^{\alpha \alpha^{\prime}}_i S^{\alpha^{\prime}}_{i,a},\\
  U^{\alpha \beta}_{ij} &\mapsto \Lambda^{\alpha \alpha^{\prime}}_i U^{\alpha^{\prime} \beta^{\prime}}_{ij} \Lambda^{\beta^{\prime} \beta}_j, &&
  \raisebox{.75\normalbaselineskip}[0pt][0pt]{$\forall\ \Lambda_i,\Lambda_j \in G,$}
    \intertext{in addition to a global $\mathrm O(3)$ symmetry}
  S^{\alpha}_{i,a} &\mapsto S^{\alpha}_{i,a^{\prime}} \Omega_{a^{\prime} a}, &&\forall\ \Omega \in \mathrm O(3).
\end{align}

The details of the gauge theory are, however, not relevant to this work (cf. Ref.~\cite{Liu16} for a comprehensive introduction to the theory).
It is utilized merely for its flexibility to produce configurations of various multipolar orders.
Specifically, since it is impossible to spontaneously break a gauge symmetry~\cite{Elitzur75}, the Hamiltonian~\eqref{eq:gauge_model} develops orders characterized by a ground-state manifold $\mathrm O(3)/G$.
Therefore, by choosing the gauge symmetry $G$, we can effectively simulate different spin orders.
For example, when $G = \mathrm O(2)$, the gauge theory recovers the Heisenberg model with general exchange interaction,
\begin{align}\label{eq:Heis}
	H_{\rm Heis} = \sum_{\langle i,j \rangle} J^{\prime}_{ab} S^{\rm n}_{i,a} S^{\rm n}_{j,b},
\end{align}
while with $G = D_{\infty h}$ it reduces to the Lebwohl-Lasher model~\cite{Lebwohl73, Lammertr93}
\begin{align}\label{eq:LL}
	H_{\rm LL} = \sum_{\langle i,j \rangle} J^{\prime\prime}_{ab} (S^{\rm n}_{i,a} S^{\rm n}_{j,b})^2.
\end{align}
We refer the reader to Ref.~\cite{Nissinen16} for a mathematical derivation of Eqs.~\eqref{eq:Heis} and \eqref{eq:LL}.

For the purpose of this paper, we will focus on a simple quadrupolar order, Eq.~\eqref{eq:LL}, a generalized quadrupolar order ($G=D_{2h}$) and a coplanar octupolar order ($G = D_{3h}$).
The input data to SVM are raw spin configurations $\mb{x} = \{S^{\alpha}_{i,a}\}$ which are generated by performing Monte Carlo simulations on the effective gauge theory.
We note that neither the existence of any order, nor the gauge fields are known to SVM. The task of SVM is to detect those orders and their order parameter without prior knowledge.

\section{Learning a single quadrupolar order} \label{sec:Dinfh}

In this section, we will take a simple quadrupolar order as an example to illustrate the basic idea of our method.
We will discuss the detection of the order and the decoding of the machine result to extract the order parameter.
Further, we will compare the performance of different training schemes; this discussion may also be applicable to other machine-learning methods for phase classification.

\subsection{Extracting the order parameter}\label{sec:Dinfh_op}

One can generate spin configurations with an emerging quadrupolar order by working with the gauge symmetry $G = D_{\infty h}$ and an anisotropic coupling where $J^{\alpha \beta}_{ab} = -J\delta_{ab}$ for $\alpha\beta = {\rm nn}$ and $J^{\alpha \beta}_{ab} = 0$ for other $\alpha \beta$'s.

The resulting model is then equivalent to the Lebwohl-Lasher model in Eq.~\eqref{eq:LL} and the $\mathbf{S}^{\rm l}$ and $\mathbf{S}^{\rm m}$ spins become irrelevant.
Thus, to simplify the exposition of our analysis, we take only the $\mathbf{S}^{\rm n}$ spins as input data to train the SVM, yielding the configuration vector $\mb{x} = \{\mb{S}^{\rm n}_i\}$. However, as will be discussed in Sec.~\ref{sec:D2h}, including $\mathbf{S}^{\rm l}$ and $\mathbf{S}^{\rm m}$ spins does not change the result.

We train the SVM successively with the multipolar kernel for increasing tensor rank, i.e. for increasing degrees $n$ of the monomials in the mapping Eq.~\eqref{eq:mapping}. As we have discussed in Ref.~\cite{Greitemann19}, when the rank is not sufficient to capture the order parameter, the SVM overfits the training data. Here, this is the case for rank $n=1$ which produces an erratic decision function, whereas we capture the order at rank $n=2$, resulting in a curve.

\begin{figure}
  \includegraphics[scale=1.0]{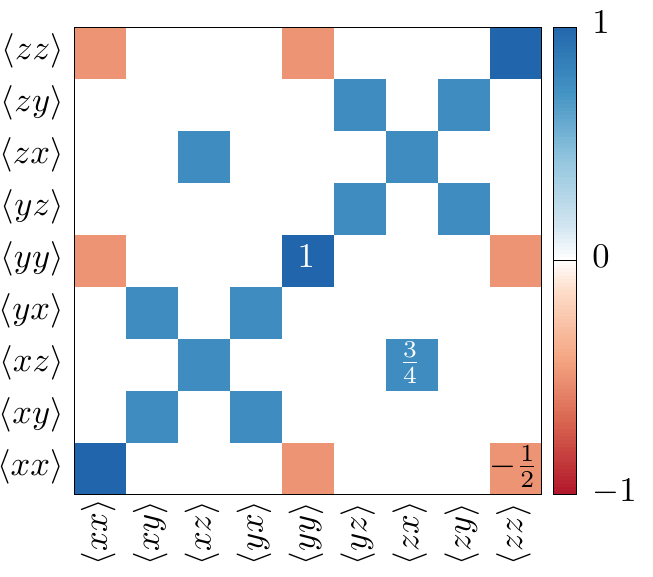}
  \caption{Coefficient matrix $C_{\mu\nu}=C_{ab,a'b'}$ for the quadrupolar order with the rank-$2$ kernel. The pattern can be decomposed according to Eq.~\eqref{eq:C_Dinfh_decomp} and leads to the uniaxial nematic tensor $Q_{ab}$.
   $28$K samples have been used for training.}
  \label{fig:Dinfh_C}
\end{figure}

By calculating the coefficient matrix (cf. Sec.~\ref{sec:kernel}), we can interpret the resulting decision function to access its analytical structure.
\begin{align}\label{eq:decision_Dinfh}
  d(\mb{x}) &= \sum_{\mu \nu} C_{\mu\nu}\phi_\mu \phi_\nu - \rho \nonumber \\
            &= \sum_{ab} \sum_{a^\prime b^\prime} C_{ab, a^\prime b^\prime}\mean{S^{\rm n}_a S^{\rm n}_b} \mean{S^{\rm n}_{a^\prime} S^{\rm n}_{b^\prime}} - \rho.
\end{align}
The coefficient matrix $C_{ab, a^\prime b^\prime}$ is shown in Fig.~\ref{fig:Dinfh_C} and encodes the contribution of the individual terms in the sum.
It can be readily decomposed into three contractions
\begin{align}
  C_{ab, a^\prime b^\prime} &= p_1 \vcenter{\hbox{\input{fig_Dinfh_C_0.tex}}} + p_2 \vcenter{\hbox{\input{fig_Dinfh_C_1.tex}}} + p_3 \vcenter{\hbox{\input{fig_Dinfh_C_2.tex}}}\label{eq:C_Dinfh_decomp}\\
  &= p_1 \delta_{a a^\prime} \delta_{b b^\prime} + p_2\delta_{a b^\prime} \delta_{b a^\prime} + p_3 \delta_{a b} \delta_{a^\prime b^\prime},
    \label{eq:C_Dinfh}
\end{align}
where we read off $p_1=p_2\approx\frac{3}{4}$ and $p_3\approx -\frac{1}{2}$.
The first two contractions are compatible with the form $\|\mathbb{O}\|_F^2 = \sum_{ab}\mathbb{O}_{ab}^2$, as they contract indices between two tensors. We shall call them `proper' contractions. The third contraction on the other hand is not, but it only produces a constant $\sum_{ab}\mean{S^{\rm n}_a S^{\rm n}_a} \mean{S^{\rm n}_b S^{\rm n}_b}=1$ independent of the configuration vector.
This `self-contraction' is an example where an internal constraint of the data (here, the normalization of the spins) allows the SVM to choose the weight $p_3$ freely in line with its optimization objective, without affecting the ability of the decision function to distinguish the phases, as pointed out in Sec.~\ref{sec:bias}.

Substituting Eq.~\eqref{eq:C_Dinfh} back to the decision function, and making use of the properties
$\mean{S^{\rm n}_a S^{\rm n}_b} = \mean{S^{\rm n}_b S^{\rm n}_a}$ and $\|\mb{S}^{\rm n} \|= 1$, we can write the decision function as the (squared) magnitude of a tensor order parameter, up to a linear rescaling:
\begin{align} \label{eq:Dinfh_Q}
	d(\mb{x}) &= \f{3}{2}\sum_{ab} \left(\mean{S^{\rm n}_a S^{\rm n}_b} - \f{1}{3} \delta_{ab}\right)^2 - \rho.
\end{align}
One identifies the (uniaxial) nematic tensor~\cite{BookdeGennes}
$Q_{ab} = \mean{S^{\rm n}_a S^{\rm n}_b} - \f{1}{3} \delta_{ab}$.
In Fig.~\ref{fig:Dinfh_decisions}(b), the rescaled decision functions are compared to the true order parameter measured from Monte Carlo simulations.

\subsection{Comparison of training schemes}\label{sec:schemes}

\begin{figure}
  \centering
  \includegraphics[scale=1.0]{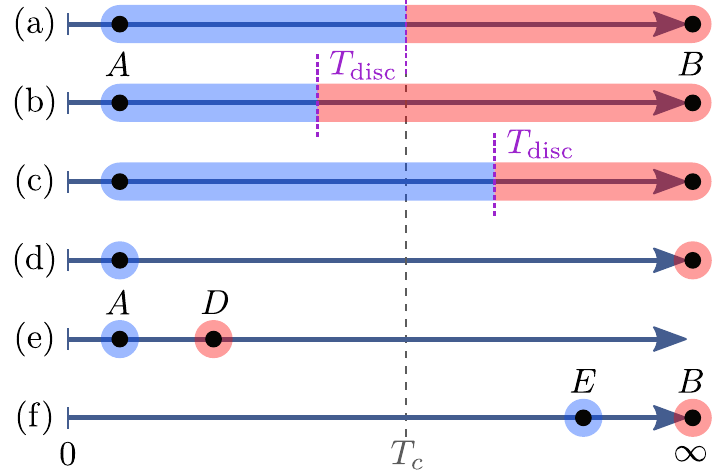}
  \caption{Graphical representation of different training schemes. Samples are taken from temperatures corresponding to the shaded regions and labeled as `ordered' (blue) or `disordered' (red). The first three schemes use a continuous distribution of temperatures from $T_A\ll T_c$ to $T_B=\infty$ and differ by the discriminatory temperature $T_\textup{disc}$ used to assign the labels, where $T_\textup{disc} = T_c = 0.52$ (a), $T_\textup{disc} = 0.4 < T_c$ (b), and $T_\textup{disc} = 0.64 > T_c$ (c) have been chosen, respectively.
  The last three schemes take samples at two discrete temperatures. (a)-(d) are discussed in Sec.~\ref{sec:schemes}; (e) and (f) in Sec.~\ref{sec:same_phase}. The concrete temperatures read as $T_A=0.2$, $T_B=\infty$, $T_D=0.25$, $T_E=2$.}
  \label{fig:Dinfh_trainings}
\end{figure}

We now examine several different schemes to generate training data for the SVM. These are illustrated in Fig.~\ref{fig:Dinfh_trainings} and labeled (a) through (f).

One can either use samples generated from a uniform distribution of temperatures crossing the phase transition, (a) through (c), or at two discrete temperatures deep inside each phase, as shown in Fig.~\ref{fig:Dinfh_trainings} (d).
When adopting the first approach, one has to set a discriminatory temperature, $T_\textup{disc}$, to label samples taken at temperatures below and above $T_\textup{disc}$. Ideally, it should coincide with the critical temperature, $T_\textup{disc}=T_c$, which we consider in scheme (a), whereas we deliberately chose $T_\textup{disc}\ne T_c$ in schemes (b) and (c) to study the effect of misclassified samples.
If the temperature distribution is discrete, the notion (and choice) of a discriminatory temperature is not necessary.
Schemes (e) and (f) are discussed in the next section.

\begin{table}
  \centering
  \renewcommand{\arraystretch}{1.2}
  \begin{tabular}{cS[table-format=2.3]|S[table-format=4.4]S[table-format=1.5]|S[table-format=4.4]S[table-format=1.5]}
    \toprule
    & & \multicolumn{2}{c|}{$\nu = 0.1$} & \multicolumn{2}{c}{$\nu = 0.4$}\\
    & $T_\textup{disc}$ & $\rho$ & $\delta$ & $\rho$ & $\delta$\\\midrule
    (a) & 0.52 & 1.012 & 0.14 & 1.001 & 0.063\\
    (b) & 0.4 & 6.83 & 1.97 & 1.003 & 0.042\\
    (c) & 0.64 & 2.32 & 0.14 & 1.0012 & 0.087\\
    (d) & n/a & 1.0012 & 0.0014 & 1.0007 & 0.0013\\
    (e) & n/a & 33.9 & 0.0090 & 25.2 & 0.0048\\
    (f) & n/a & 0.15 & 1.78 & 6.1 & 0.41\\\bottomrule
  \end{tabular}
  \caption{The SVM bias parameter $\rho$ and the deviation $\delta$ from the true uniaxial nematic order parameter are given for 28K samples generated according to the training schemes (a)-(f) and for weak ($\nu=0.1$) and strong ($\nu=0.4$) regularization. $\rho\approx 1$ indicates a phase transition was captured.}
  \label{tab:rho_delta}
\end{table}

In all cases, we measured the decision function and examined the  $C_{\mu \nu}$ matrix as before.
As anticipated, the coefficient matrix exhibits the same pattern as shown in Fig.~\ref{fig:Dinfh_C} for continuous (a) and discrete (d) temperature distributions, meaning the physical order parameter is captured regardless.

To further quantify the performance of these training schemes, we introduce a deviation metric,
\begin{align}
	\delta \coloneqq \frac{\|\mb{C} -
  \tilde{\mb{C}}\|_F}{\|\tilde{\mb{C}}\|_F} \geq 0,
\end{align}
which measures the element-wise discrepancy between the learned $C_{\mu \nu}$ and the theoretical one, $\tilde{C}_{\mu \nu}$. We note that this metric is rather sensitive and a value of $\delta=0.1$ already constitutes a rather good result.

Table~\ref{tab:rho_delta} shows results for the various training schemes and for weak ($\nu=0.1$) and strong ($\nu=0.4$) regularization. Aside from the deviation $\delta$, we tabulate the bias $\rho$ which is accessible as part of the optimization result and we expect it to attain a value of $1$ if the SVM learned a physical order parameter for reasons laid out in Sec.~\ref{sec:bias}.

First, we observe that the order parameters extracted using strong regularization are always better than those obtained using weaker regularization. This is consistent with our previous findings for high-rank order parameter tensors in Ref.~\cite{Greitemann19}.

We also observe that training set (d) in fact yields the best results, even compared to set (a) where the accurate critical temperature was used to discriminate the phases. Thus, SVM seems to work best if the training data exhibit the characteristics of the ordered phase most pronounced. It does not benefit from training data in the vicinity of the critical point.

However the continuous training sets (a) through (c) allow for a validation of the tentative phase diagram. Whereas in (a) the correct order parameter is learned for both degrees of regularization, the misclassification of the samples with temperatures between $T_\textup{disc}$ and $T_c$ in sets (b) and (c) induces significant deviations from the true order parameter at weak regularization which is compensated for at stronger regularization. If continuous training is used in this way to verify the existence of a transition at $T_\textup{disc}$, a relatively weak regularization may be desired.

We also note the correlation between a small deviation $\delta$ and a bias $\rho$ close to one in all cases for (a) through (d). This is consistent with our expectation of $\rho=1$ for a physical order parameter and enables us to use the bias to gauge the quality of the learned order parameter without examining its analytical structure.
In case (b), samples from the ordered phase are wrongly labeled as disordered and push the SVM margin boundary away from the manifold of truly disordered samples which both induces a bias $\rho\gg 1$ and deforms the decision boundary, resulting in a faulty order parameter, $\delta\gg 0$. Likewise, in the opposite case (c), which falsely classifies disordered samples as ordered ones, the margin is forcibly kept very narrow which makes it prone to overfitting.

\begin{figure}
  \centering
  \includegraphics[scale=1.0]{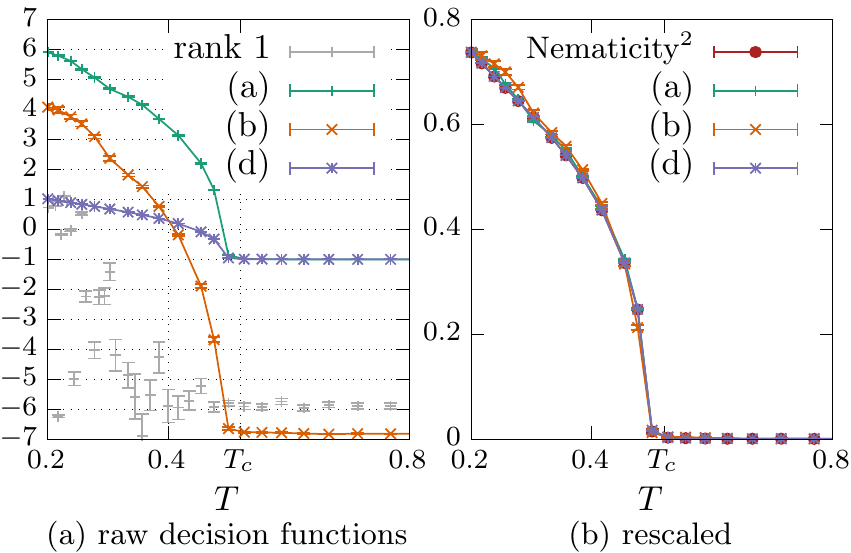}
  \caption{The decision function, obtained from the earlier training with regularization $\nu=0.1$, is measured as an observable in Monte Carlo simulations at temperatures spanning the phase transition. The left panel shows the numerical values of the decision functions as obtained from SVM. When training the SVM with the rank-1 kernel (shown in gray), the quadrupolar order cannot be captured, resulting in an erratic decision function. The curves (a), (b), and (d) are trained with the rank-2 kernel and correspond to the eponymous training schemes (cf. Fig.~\ref{fig:Dinfh_trainings}). The zero crossing of the decision functions marks the decision boundary, whereas the constant value for $T>T_c$ is given by $-\rho$ (cf. Tab.~\ref{tab:rho_delta}). The right panel shows the same curves, but shifted up by $\rho$, and are seen to match the square of the uniaxial nematicity after being rescaled.}
  \label{fig:Dinfh_decisions}
\end{figure}

In the left panel of Fig.~\ref{fig:Dinfh_decisions}, we show the ``raw'' decision functions without any rescaling. Note that the constant value attained in the disordered phase corresponds to $-\rho$. The absolute scale of the decision function is related to the inverse width of the SVM margin. For the continuous sampling schemes, the margin is more narrow as it is constrained by samples close to the transition. In the discrete scheme (d), the decision function gives values between $-1$ and $1$, corresponding to the extreme temperatures it was trained at.
Further, we point out that the zero crossing of the raw decision function, which determines the label of a testing sample, approximately reproduces the discriminatory temperatures used in the continuous training schemes (a) and (b), whereas the discrete scheme (d) is not trained to classify samples close to the transition.

While these features of the `raw' decision function can provide a deeper understanding of the workings of the SVM, they are physically irrelevant and we therefore shift the decision functions by $\rho$ to obtain a value of zero in the disordered phase and rescale them to match the true nematicity. As can be seen from the right panel in Fig.~\ref{fig:Dinfh_decisions}, the result is very close in all three cases, even for scheme (b).


\subsection{Training in the same phase}\label{sec:same_phase}

We now examine two more training schemes where the samples are taken at two discrete temperatures which are both in the ordered phase (e) or disordered phase (f). We do so to assess the capability to distinguish these cases where no phase transition takes place from those where it does, (a)-(d).
This can be physically relevant in situations where the phase boundary is not known, source data is limited, or when multiple order parameters are involved, while not all of them experience a phase transition.

Interestingly, when both training temperatures lie within the ordered phase (e), the SVM still manages to learn the order parameter, as is apparent from the very small deviations $\delta$ in Tab.~\ref{tab:rho_delta}.
This can be explained by the fact that the magnitude of the true order parameter is not constant within the ordered phase, but increases in value as one moves deeper inside the phase. The true order parameter is still the best decision function the SVM can learn in this situation to distinguish samples from the two temperatures. Note that this enables us to measure the decision function at a range of temperatures which can even exceed the training temperatures and read off the factual transition point from the decision function.

Despite the excellent $\delta$ in scenario (e), the corresponding biases $\rho$ are far away from unity. This is no surprise as the argument put forth in Sec.~\ref{sec:bias} relied on the properties of disordered samples. Thus, the bias enables us to distinguish this situation from case (d) where a transition takes place between the training temperatures.

We get a different picture when we attempt to train the SVM with samples from two temperatures in the disordered phase (f). The magnitude of the true tensor order parameter is constant (zero) throughout the disordered phase which makes it unsuitable to distinguish samples from the two temperatures. Thus, SVM does not learn the correct order parameter, as is manifest from the large value of $\delta$ in Table~\ref{tab:rho_delta}, but instead overfits the training data in an attempt to construct a better decision function. As a further indication of overfitting, the decision function is not reproducible. Likewise, also the bias $\rho$ is fluctuating and in general will not be close to one.


\section{Learning multiple orders}\label{sec:biaxial}
If a spin or orbital system of interest develops an order which needs to be characterized by some axial point group, in general, it may require more than one (multipolar) order parameter.
For example, a generalized quadrupolar order is defined by two rank-2 tensors.
In this section, we demonstrate the ability of our kernel Eq.~\eqref{eq:kernel} to detect multiple orders simultaneously.
We consider situations where the orderings occur at the same rank and at different ranks and assume here that the topology of the phase diagram is approximately known. In Sec.~\ref{sec:mapping_PD}, we propose a scheme that yields a phase diagram in situations where such information is absent.

\subsection{Order parameters of the same rank} \label{sec:D2h}
A biaxial order $D_{2h}$ is considered.
The training samples are prepared by choosing the gauge symmetry $G=D_{2h}$ and the exchange coupling
$J^{\rm ll}_{ab} = J^{\rm mm}_{ab}= -J_1 \delta_{ab}$,
$J^{\rm nn}_{ab} = -J_3 \delta_{ab}$ in Eq.~\eqref{eq:gauge_model}. The gauge model is then equivalent to the Straley model of generalized quadrupolar orders~\cite{Straley74}.
$J_1$ and $J_3$ are used as tuning parameters for convenience.

\begin{figure}
  \centering
  \includegraphics[scale=1.0]{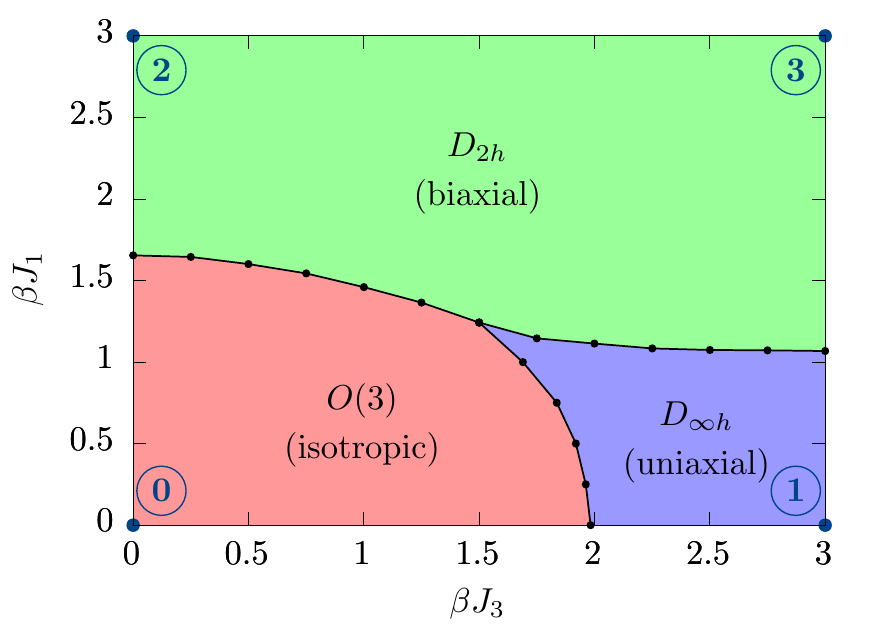}
  \caption{The phase diagram for the generalized quadrupolar order (cf. Ref.~\cite{Liu17}). The points labeled \protect\circled{0} through \protect\circled{3} are used to generate training samples.}
  \label{fig:D2h_PD}
\end{figure}

The associated phase diagram is shown in Fig.~\ref{fig:D2h_PD}. It requires at least two order parameters to characterize the three phases and their transitions.
If the two orders are weakly coupled, they will experience (dis-)ordering separately with one of them being irrelevant in the corresponding phase transition.
However, if the coupling between them is strong, they may develop ordering in a single phase transition and need to be taken into account simultaneously. This is the case in the transition between the biaxial phase and the isotropic phase.

Given the topology of the phase diagram, without knowledge of the exact phase boundary, we can train the SVM with configurations sampled at points deep in each phase.
Here, we took samples at four points, \circled{0}--\circled{3}, corresponding to the corners of the phase diagram shown in Fig.~\ref{fig:D2h_PD}.

\begin{table}
  \centering
  \renewcommand{\arraystretch}{1.2}
  \begin{tabular}{rcl|S[table-format=2.5]|S[table-format=2.5]}
    \toprule
    &&& \multicolumn{1}{c|}{$\nu = 0.1$} & \multicolumn{1}{c}{$\nu = 0.5$}\\
    &&& $\rho$ & $\rho$\\\midrule
    \circled{0} & / & \circled{1}\ \ & 1.0017 & 1.0007\\
    \circled{0} & / & \circled{2}\ \ & 1.0013 & 1.0007\\
    \circled{0} & / & \circled{3}\ \ & 1.0010 & 1.0006\\
    \circled{1} & / & \circled{2}\ \ & 0.972  & 0.9995\\
    \circled{1} & / & \circled{3}\ \ & 1.27   & 1.34  \\
    \circled{2} & / & \circled{3}\ \ & 8.05   & 7.50  \\
    \bottomrule
  \end{tabular}
  \caption{The value of the bias parameter in learning the phases of the generalized quadrupolar order.
  Training samples are collected from the corresponding regimes in Fig.~\ref{fig:D2h_PD}) and labeled accordingly.
  This results in six classifiers, each distinguishing between a pair of those four labels.
  A total of 25k samples have been used in the training, and results from weak ($\nu=0.1$) and strong ($\nu=0.5$) regularization are shown for comparison.
  Note that for the classifier between \protect\circled{2} and \protect\circled{3}, we observe $\rho\gg 1$, indicating that no phase transition takes place between those points.}
  \label{tab:D2h_4way_rho}
\end{table}

\begin{table}
  \centering
  \renewcommand{\arraystretch}{1.2}
  \begin{tabular}{rcl|S[table-format=2.5]|S[table-format=2.5]|c}
    \toprule
    &&& \multicolumn{1}{c|}{$\nu = 0.1$} & \multicolumn{1}{c|}{$\nu = 0.5$} &\\
    &&& $\rho$ & $\rho$ & \ \ Fig.\ \ \\\midrule
    $\mathrm O(3)$ & / & $D_{\infty h}$ & 1.0017 & 1.0007 & \ref{fig:D2h_C_U}\\
    $D_{\infty h}$ & / & $D_{2h}$ & 0.985 & 1.0004 & \ref{fig:D2h_C_B}a\\
    $\mathrm O(3)$ & / & $D_{2h}$ & 1.0011 & 1.0005 & \ref{fig:D2h_C_B}b\\\bottomrule
  \end{tabular}
  \caption{Same as Tab.~\ref{tab:D2h_4way_rho}, but now the data sampled at points \protect\circled{2} and \protect\circled{3} are assigned the same label `$D_{2h}$'; correspondingly \protect\circled{0} is labeled as `$\mathrm O(3)$' and \protect\circled{1} as `$D_{\infty h}$'. In all cases, the biases are close to one, indicating that the labeling indeed represents the phases. The last column refers to the figures showing the (block structure of the) coefficient matrix corresponding to that classifier.}
  \label{tab:D2h_deep_rho}
\end{table}

With the multipolar kernel at rank 2, we now perform multiclassification as described in Sec.~\ref{sec:multi} where we labeled the samples with $M=4$ labels corresponding to the points \circled{0}--\circled{3}. This yields six classifiers. Their respective biases are tabulated in Table~\ref{tab:D2h_4way_rho}. We can note straight away that the classifier distinguishing points \circled{2} and \circled{3} does not correspond to a phase transition as its bias is far away from one. This is consistent with our knowledge that these points are indeed both in the biaxial phase. The bias of the classifier between \circled{1} and \circled{3} is also somewhat elevated, so it is not clear that a phase transition takes place. However, after merging the labels \circled{2} and \circled{3} and repeating the multiclassification with $M=3$ distinct labels (which now correspond to the phases), we find that the bias of the corresponding classifier is closer to one (cf. Tab.~\ref{tab:D2h_deep_rho}), affirming that a phase transition is present. The larger deviation from one in this case can be explained by the fact that the uniaxial order is developed in both phases whereas the magnitude of the ordering is larger in the biaxial phase.


\begin{figure}
  \centering
  \includegraphics[scale=1.0]{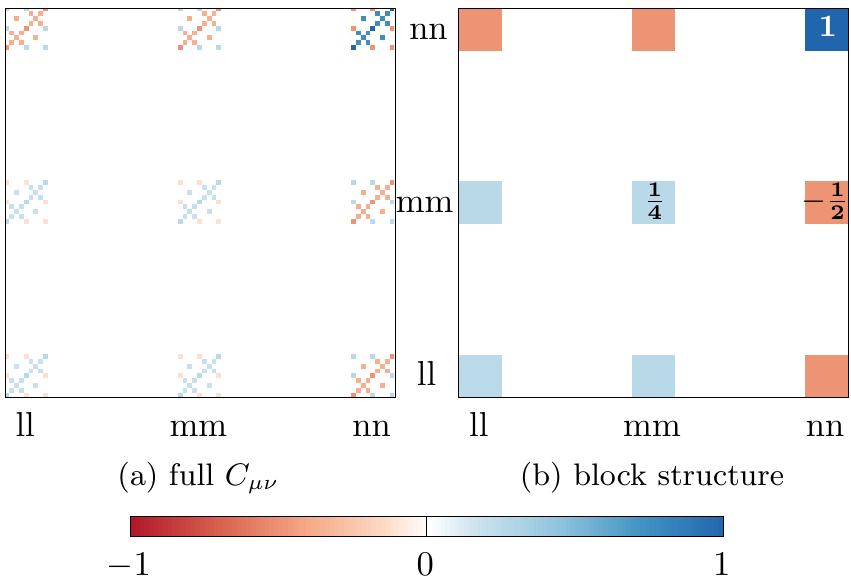}
  \caption{The coefficient matrix distinguishing the isotropic ($\mathrm O(3)$) and uniaxial ($D_{\infty h}$) phase of the generalized quadrupolar order. The left panel displays the full matrix $C_{\mu\nu}$ where the multi-indices $\mu$ and $\nu$ are sorted lexicographically and where the color indices are more significant. The resulting block structure is shown in the right panel. The block weights have been obtained by summing all matrix elements within each block over their component indices and normalizing to the top-right $({\rm nn,nn})$ block.
  The interpretation leads to an alternative expression of the nematic $Q_{ab}$ tensor, given in Eq.~\eqref{eq:Dinfh_op_lm}.
  The pattern exhibited within each block is identical to Fig.~\ref{fig:Dinfh_C}.}
  \label{fig:D2h_C_U}
\end{figure}

We now proceed with the analysis of the three coefficient matrices obtained from the latter classification with respect to three labels.
Let us first revisit the nematic order which is now responsible for the isotropic-to-uniaxial transition in the regime of small $J_1$ in Fig.~\ref{fig:D2h_PD}.
The corresponding $C_{\mu\nu}$ matrix is shown in Fig.~\ref{fig:D2h_C_U}(a).
With $\mb{S}^{\rm l}$ and $\mb{S}^{\rm m}$ degrees of freedom now included, it contains $81\times 81$ elements, defining contractions between  $\phi_\mu = \mean{S^{\alpha_1}_{a_1} S^{\alpha_2}_{a_2}}$ and $\phi_\nu = \mean{S^{\beta_1}_{b_1} S^{\beta_2}_{b_2}}$.
However, when arranging the multi-indices such that color indices are more significant than component indices (i.e. $\mu=(\alpha_1,\alpha_2;a_1,a_2)$), it can be divided into a $9$-by-$9$ block structure in terms of the color indices $(\alpha_1 \alpha_2), (\beta_1 \beta_2)$, corresponding to the $9$ basis tensors in Eq.~\eqref{eq:O_tensor}.
Each block in fact features the same pattern of Fig.~\ref{fig:Dinfh_C} up to a weight factor. In Sec.~\ref{sec:Dinfh_op}, its relation to the order parameter tensor was explained. In the following analysis, we can ignore the internal pattern of the blocks entirely and instead infer the coordinates of the order parameter tensor in the nine-dimensional tensor space from the relative weight factors of those blocks.
The weight of a block may be obtained by a sum over all elements within the block as $\sum_{\mb{ab}}C^{\bds{\alpha\beta}}_{\mb{ab}}$, or alternatively by the Frobenius norm as $\sqrt{\sum_{\mb{ab}}(C^{\bds{\alpha\beta}}_{\mb{ab}})^2}$.
We have used both definitions and verified that they yield the same weights up to a sign.
The first definition preserves the sign of the block weights, so we adopted it.
With this definition, normalized to the most pronounced $({\rm nn, nn})$ block in the top-right corner in Fig.~\ref{fig:D2h_C_U}(a), the weights of the $({\rm ll, ll})$ and $({\rm mm, mm})$ blocks turn out to be $\f{1}{4}$, while those of the $({\rm nn, ll})$ and $({\rm nn, mm})$ blocks are $-\f{1}{2}$ [cf. Fig.~\ref{fig:D2h_C_U}(b)].
This gives an order parameter of the form
\begin{align}\label{eq:Dinfh_op_lm}
	\mbb{O}_{\rm uni} = \mb{S}^{\rm n} \otimes \mb{S}^{\rm n} - \f{1}{2}\mb{S}^{\rm m} \otimes \mb{S}^{\rm m} - \f{1}{2} \mb{S}^{\rm l} \otimes \mb{S}^{\rm l}.
\end{align}
Using the relation
 $\sum_{\alpha = {\rm l,m,n}}\mb{S}^{\rm \alpha} \otimes \mb{S}^{\rm \alpha} = \id$,
  we recover the nematic tensor of Eq.~\eqref{eq:Dinfh_Q}:
\begin{align} \label{eq:Dinfh_op}
	\mbb{O}_{\rm uni} = \f{3}{2}(\mb{S}^{\rm n} \otimes \mb{S}^{\rm n} -\f{1}{3} \id).
\end{align}
Thus, to extract the order parameter, it is sufficient to rely on the block structure of $C_{\mu\nu}$, rather than the full matrix.
This simplifies the interpretation significantly, especially for learning high-rank \cite{Greitemann19} and multiple simultaneous orders.

\begin{figure}
  \centering
  \includegraphics[scale=1.0]{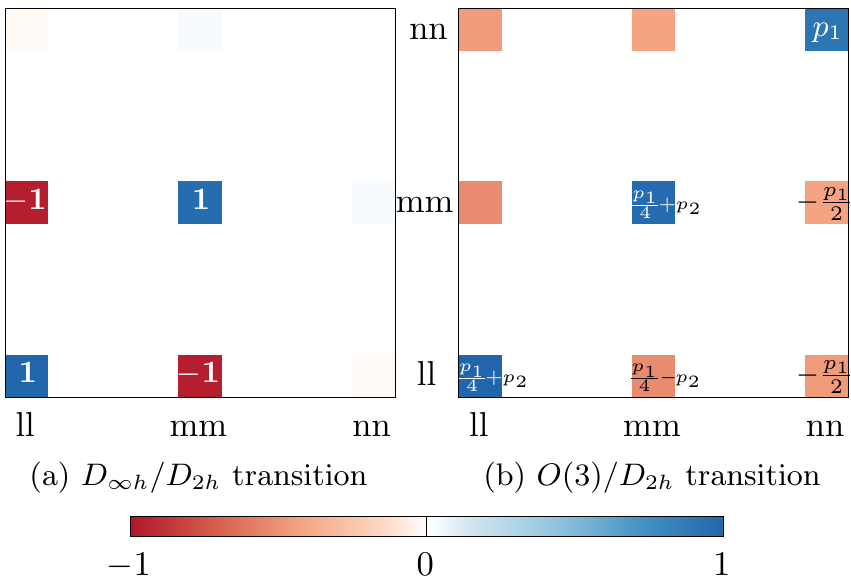}
  \caption{The block structures of the $C_{\mu\nu}$ matrix of the remaining two decision functions of the generalized quadrupolar order,
   (a) the uniaxial ($D_{\infty h}$) and biaxial ($D_{2h}$) phases, and (b) the isotropic [$\mathrm O(3)$] and biaxial ($D_{2h}$) phases. The latter is a superposition of the patterns exhibited by the other two decision functions, shown in Figs.~\ref{fig:D2h_C_U}(b) and \ref{fig:D2h_C_B}(a).
   The block weights are not fixed, but rather the coefficients $p_1$ and $p_2$ vary with the relative strength of the two orderings.}
  \label{fig:D2h_C_B}
\end{figure}

In Fig.~\ref{fig:D2h_C_B}, we show the block structure of the coefficient matrix for the other two classifications.  The direct sum is used to compute the block weights.
Figure~\ref{fig:D2h_C_B}(a) is the weight matrix trained with data sampled deep inside the uniaxial and biaxial phases.
One immediately realizes a tensor of the form
\begin{align} \label{eq:op_biaxial}
	\mbb{O}^{(D_{2h})}_{\rm bi} = \mb{S}^{\rm l} \otimes\mb{S}^{\rm l} - \mb{S}^{\rm m} \otimes \mb{S}^{\rm m},
\end{align}
which is indeed the biaxial order parameter describing the uniaxial-to-biaxial phase transition~\cite{BookdeGennes}.

Fig.~\ref{fig:D2h_C_B}(b) is the weight matrix trained with samples from the isotropic and the biaxial phases.
It is a superposition of that in Figs.~\ref{fig:D2h_C_U}(b) and \ref{fig:D2h_C_B}(a), with the weight denoted by $p_1$ and $p_2$.
Its interpretation then leads to a set of two order parameters, $\{ \mbb{O}_{\rm uni}, \mbb{O}^{(D_{2h})}_{\rm bi} \}$, which is indeed the order parameter set that uniquely defines the $D_{2h}$ biaxial phase.
[Fig.~\ref{fig:D2h_C_B}(b) can also be viewed as a superposition of three order parameters,
$\mb{S}^{\rm l} \otimes\mb{S}^{\rm l} - \mb{S}^{\rm m} \otimes \mb{S}^{\rm m}$,
$\mb{S}^{\rm m} \otimes\mb{S}^{\rm m} - \mb{S}^{\rm n} \otimes \mb{S}^{\rm n}$
and  $\mb{S}^{\rm n} \otimes\mb{S}^{\rm n} - \mb{S}^{\rm l} \otimes \mb{S}^{\rm l}$,
where two of them are independent. This is just an equivalent expression of the same ordering.]

We find that the weights of $p_1$ and $p_2$ vary as we change the points in phase diagram from which we sample. This does not come unexpected, as the cooccurrence and the expression of the two orders are protected by symmetry (of their ground-state manifold), so their relative ratio depends on the microscopic coupling strengths (the value of $J_1$ and $J_3$ in this case).

This further emphasizes the importance of interpretability. In a physical problem, we typically treat different order parameters as individual quantities and measure them separately.
However, a machine may combine them into a single numerical classifier.
Therefore, a machine trained by samples with certain relative ratios of those order parameters may not be optimal or even misleading when being applied to samples where those ratios vary.
In order to make correct predictions, one may need to be able to identify and isolate each order parameter from the machine result.
This is, however, not an issue for our kernel method (or in general for an interpretable machine). The occurrence of multiple orders only leads to a {\it linear} superposition of the pattern of each single order, even though those orders may be strongly coupled.

\subsection{Order parameters of different ranks} \label{sec:D3h}

The quadrupolar and octupolar orders in classical kagome antiferromagnets provide examples for spin systems which simultaneously develop multipolar orders of different ranks.
Indeed, the quadrupolar order in Eq.~\eqref{eq:Dinfh_op} is compatible with all dihedral symmetries $D_{n}, D_{nh},  D_{nd}$, and axial symmetries $S_{2n}, C_{nh}$.
That is, if the ground-state manifold of a spin order cannot be ascertained theoretically, we should not immediately conclude a uniaxial phase after observing a quadrupolar order.
This is also true for dipolar (rank-1) orders which are compatible with axial symmetries $C_n$ and $C_{nv}$, while the Heisenberg and N{\'e}el magnetization are just limiting cases for $n \rightarrow \infty$, $\mathrm O(2) = C_{\infty v}$.
Therefore, to correctly characterize a multipolar phase, in principle we need to identify all relevant orders.

The kernel~\eqref{eq:kernel} handles these situations in a straightforward manner. We simply train the SVM with kernels at different ranks $n$ separately.
Moreover, considering the crystallographic background, it is sufficient to set an upper bound $n\le 6$.
Additional complexity may arise in distinguishing nontrivial high-rank orders from responses which are already captured by order parameters of lower rank.
For example, if we measure the quadrupolar order parameter $Q_{ab} = S_a S_b - \frac{1}{3} \delta_{ab}$ in a ferromagnetic phase, it will also show a finite response. This quadrupolar order is nevertheless trivial since it is completely captured by a dipolar order, namely, the magnetization.
However, this is easily resolved if the machine is interpretable, as is the case for our kernel.
We can extract the analytical expression of the learned order parameters at different ranks, and identify those without lower-rank origins.

To be concrete, we take a simultaneous occurrence of quadrupolar and octupolar orders as an example.
This can be realized by choosing $G = D_{3h}$ in the gauge theory Eq.~\eqref{eq:gauge_model}.
Analogously to the previous section, the training samples are collected from deep within the ordered $D_{3h}$ phase, the uniaxial $D_{\infty h}$ phase, and the isotropic $\mathrm O(3)$ phase.

We then carry out the SVM optimization with kernels of ranks 1 through 6.
The biases of the resulting classifiers are tabulated in Table~\ref{tab:D3h_deep_rho}. We see that rank 1 is insufficient to capture any order.
At rank 2, we find an order parameter describing both the isotropic-to-uniaxial and isotropic-to-biaxial transitions, whereas no order parameter describing the uniaxial-to-biaxial transition could be found. The coefficient matrix $C_{\mu\nu}$ is identical to that of Fig.~\ref{fig:D2h_C_U}, indicating a quadrupolar order has been captured.
At rank 3, matters are reversed and we capture the uniaxial-to-biaxial transition but not the isotropic-to-uniaxial one. The isotropic-to-biaxial transition exhibits a coexistence of both the quadrupolar and the rank-3 order parameter. At higher ranks, we do not find new nontrivial order parameters but those that we do find are functions of the lower-rank ones. At rank 4, we essentially learn the square of the quadrupolar order parameter. At ranks 5 and 6, the tensor can be constructed from both of the lower-rank ones. Rather than relying on the bias criterion to determine if an order parameter was found, one may also measure the decision functions and discard them if they exhibit erratic behavior, similarly to how rank 1 was ruled out as an order parameter in Sec.~\ref{sec:Dinfh_op}. This approach leads to the same conclusions.

\begin{table}
  \centering
  \renewcommand{\arraystretch}{1.2}
  \begin{tabular}{c|S[table-format=2.5]|S[table-format=2.5]|S[table-format=2.5]}
    \toprule
    & \ {$\mathrm O(3) / D_{\infty h}$}\ & \ {$D_{\infty h} / D_{3h}$}\  & \ {$\mathrm O(3) / D_{3h}$}\ \\
    \ {rank}\ \ & $\rho$ & $\rho$ & $\rho$\\\midrule
    1 & 0.134  & 2.95   & 2.61  \\
    2 & 1.0019 & 5.87   & 1.0018\\
    3 & 4.10   & 1.0012 & 1.0011\\
    4 & 1.0029 & 0.617  & 1.0025\\
    5 & 0.981  & 1.0012 & 1.0011\\
    6 & 1.0018 & 1.087  & 1.0016\\
    \bottomrule
  \end{tabular}
  \caption{The biases of the three SVM classifiers discerning the isotropic [$\mathrm O(3)$], uniaxial ($D_{\infty h}$) and octupolar ($D_{3h}$) phases.
  Decision functions are monitored for ranks $1$ through $6$ of the kernel. As before, $\rho\approx 1$ indicates that an order parameter of that rank could be learned. This is the case for the uniaxial order at ranks 2, 4, 5, and 6, and for the biaxial order at ranks 3, 5, and 6. In both cases, the higher-rank ones are trivial functions of their lowest-rank representation. The transition between the isotropic and biaxial phase in the last column experiences simultaneous uniaxial and biaxial ordering and will thus learn an order parameter for any rank $n\ge 2$.}
  \label{tab:D3h_deep_rho}
\end{table}

\begin{figure}
  \centering
  \includegraphics[scale=1.0]{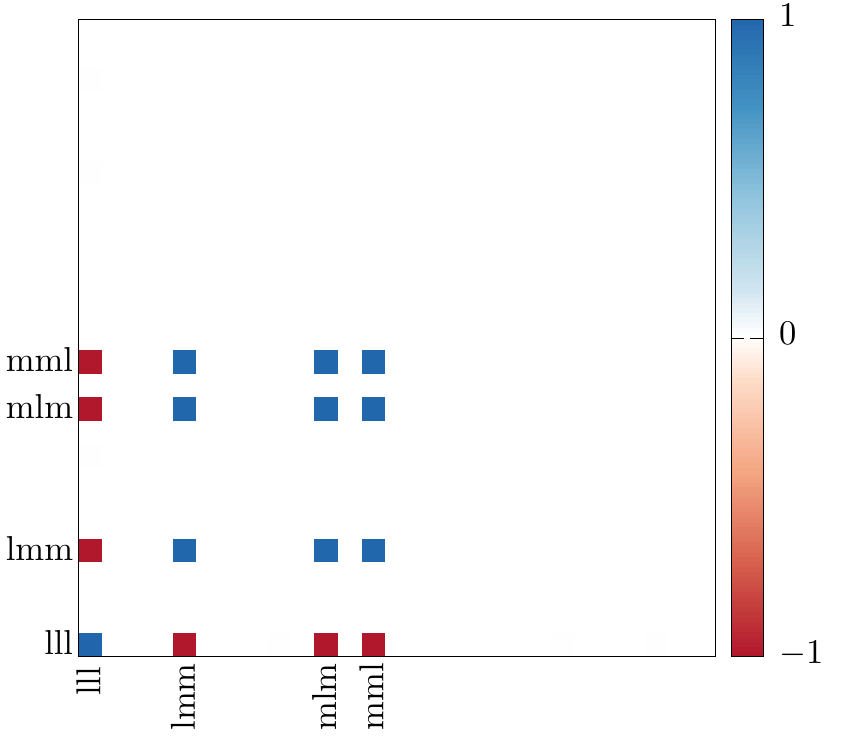}
  \caption{The block structure of the $C_{\mu\nu}$ matrix learned by the rank-$3$ kernel, distinguishing the octupolar phase ($D_{3h}$) from the isotropic [$\mathrm O(3)$] and the uniaxial phase ($D_{\infty h}$). The color indices are sorted lexicographically and those of the nonvanishing blocks are explicitly given.
  The interpretation of this pattern gives the rank-$3$ octupolar ordering tensor in Eq.~\eqref{eq:D3h_op_2}.}
  \label{fig:D3h_C}
\end{figure}

\begin{figure}
  \centering
  \includegraphics[scale=1.0]{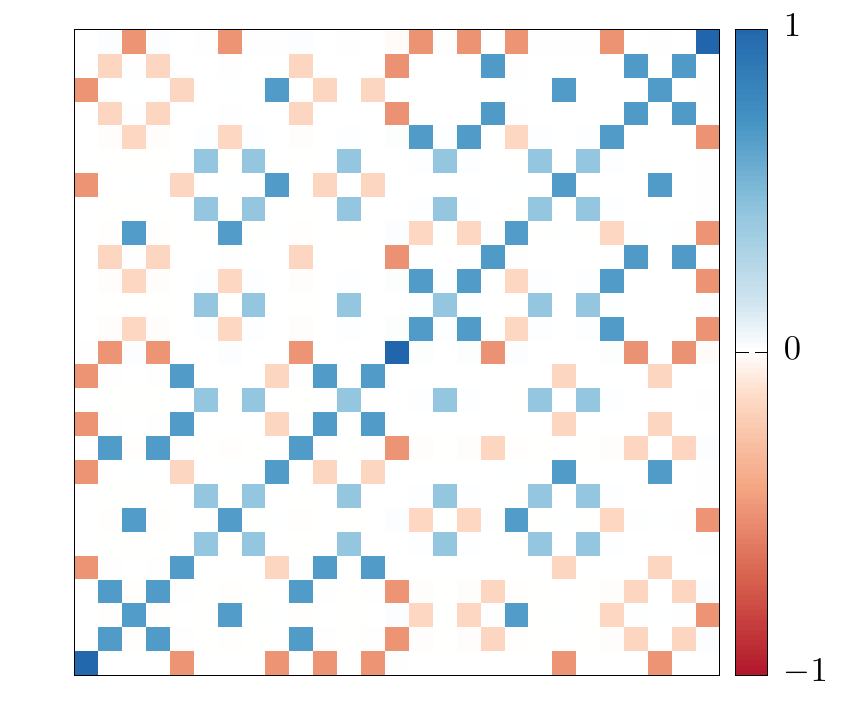}
  \caption{Internal structure of the $({\rm lll,lll})$ block of the coefficient matrix in Fig.~\ref{fig:D3h_C}. The axes are labeled by lexicographically sorted triplets of component indices ($xxx$, $xxy$, \ldots).
  This pattern corresponds to the expression Eq.~\eqref{eq:C_D3h_lll} and in turn gives the one-spin form of the octupolar ordering tensor in Eq.~\eqref{eq:D3h_op_lll}.}
  \label{fig:D3h_C_lll}
\end{figure}

The coefficient matrix for the rank-3 order is shown in Fig.~\ref{fig:D3h_C}. For conciseness, we omit the full matrix and only show its block structure.
We infer the contraction of two tensors of the form
\begin{align}\label{eq:D3h_op_2}
	\mathbb{O}^{(D_{3h})}_{\rm bi} =&\ \mb{S}^{\rm l} \otimes \mb{S}^{\rm l} \otimes\mb{S}^{\rm l} - \mb{S}^{\rm l} \otimes \mb{S}^{\rm m} \otimes\mb{S}^{\rm m} \nonumber \\
	&- \mb{S}^{\rm m} \otimes \mb{S}^{\rm l} \otimes\mb{S}^{\rm m} - \mb{S}^{\rm m} \otimes \mb{S}^{\rm m} \otimes\mb{S}^{\rm l},
\end{align}
which is the octupolar order parameter that was also found in classical kagome antiferromagnets~\cite{Zhitomirsky08}.
It is a two-spin form of the $D_{3h}$ octupolar order. However, for a coplanar order, the $\mb{S}^{\rm m}$ spin is arbitrary in the sense that we can always introduce another spin orthogonal to $\mb{S}^{\rm l}$. Therefore, the octupolar order may also be defined by the $\mb{S}^{\rm l}$ spins alone.
To obtain the one-spin representation, we simply need to examine elements inside the $({\rm lll, lll})$ block of the $C_{\mu\nu}$ matrix, shown in Fig.~\ref{fig:D3h_C_lll}.

This is similar to the discussion of the quadrupolar order in Sec.~\ref{sec:Dinfh_op}, but involves more contractions owing to the higher rank.
Still, these contractions and their weight can be readily inferred,
\begin{align} \label{eq:C_D3h_lll}
	C^{\rm lll,lll}_{abc,a^{\prime}b^{\prime}c^{\prime}} = & \ p_1 \big(\delta_{aa^{\prime}} \delta_{bb^{\prime}} \delta_{cc^{\prime}} +
	\textrm{permutations of } \{a^{\prime}b^{\prime}c^{\prime}\}
 	\big) \nonumber \\
	& + p_0 \big(\delta_{aa^{\prime}} \delta_{bc}\delta_{b^{\prime}c^{\prime}}
	+ \textrm{other self-contractions}\big),
\end{align}
where $p_1$ and $p_0$ denote the weight of proper and self-contractions, respectively, and a relation $p_1 = -\f{5}{2}p_0$ is found up to numeric precision.

As $\mb{S}^{\rm l}\bigr.^{\otimes 3}$ is symmetric, the six proper contractions in Eq.~\eqref{eq:C_D3h_lll} are equivalent if we substitute them to the decision function [as in Eq.~\eqref{eq:Dinfh_Q}].
Similarly, the nine self-contractions in this case can also be grouped into three equivalent classes.
Thus, Eq.~\eqref{eq:C_D3h_lll} can effectively be expressed as
\begin{align}
	C^{\rm lll,lll}_{abc,a^{\prime}b^{\prime}c^{\prime}} = &\ \delta_{aa^{\prime}} \delta_{bb^{\prime}} \delta_{cc^{\prime}} - \f{1}{5} \delta_{aa^{\prime}} \delta_{bc} \delta_{b^{\prime}c^{\prime}} \nonumber \\
	& -\f{1}{5} \delta_{ac} \delta_{bb^{\prime}} \delta_{a^{\prime}c^{\prime}}
	-\f{1}{5} \delta_{ab} \delta_{a^{\prime}b^{\prime}} \delta_{cc^{\prime}},
\end{align}
whose interpretation leads to a tensor
\begin{align} \label{eq:D3h_op_lll}
	T_{abc} =  S^{\rm l}_aS^{\rm l}_b S^{\rm l}_c - \f{1}{5} S^{\rm l}_a \delta_{bc} - \f{1}{5} S^{\rm l}_b \delta_{ca} - \f{1}{5} S^{\rm l}_c \delta_{ab}.
\end{align}
This is exactly the one-spin form of the octupolar order given in Ref.~\cite{Zhitomirsky08}.

\section{Exploring the phase diagram} \label{sec:mapping_PD}

In the preceding two sections, we have demonstrated that the bias $\rho$ can in fact be used to ascertain whether or not the labeling of the training samples is consistent with a phase transition. Motivated by this, we embrace this criterion as the central element of a learning scheme that relegates the supervision aspect to its weakest possible sense.

In fact, we will not assume any prior knowledge of the topology of the phase diagram. We impose a uniform grid on the parameter space and sample configurations at each of those grid points, labeling the samples accordingly. This sets up a multiclassification problem where the number of classes equals the number of grid points, which we shall call $M$. Applying SVM to the latter yields $M(M-1)/2$ classifiers whose biases may be used to consider their respective grid points as belonging to the same phase or not. Specifically, we can build an undirected graph whose vertices correspond to the grid points and are connected by an edge if the bias $\rho$ of the corresponding classifier exceeds a given threshold value $\rho_c$. Larger values of $\rho_c$ will result in graphs with less false-positive edges at the cost of more false-negative (missing) edges. Since we are considering all pairings of grid points, the graph contains a lot of redundant information, so false-negatives are more easily compensated for. We discuss the choice of $\rho_c$ below.

\begin{figure}
  \centering
  \includegraphics[scale=1.0]{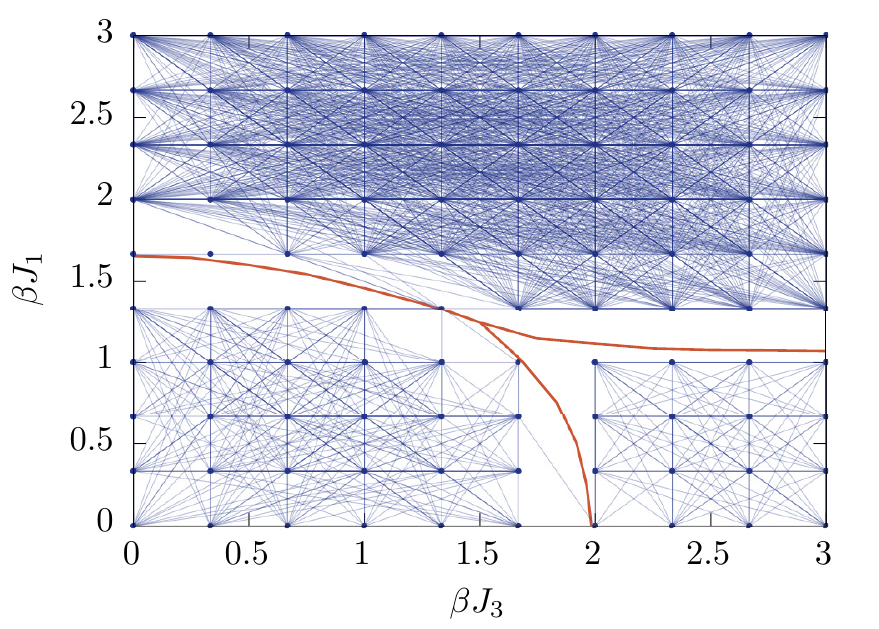}
  \caption{Graph representation of the relation between points on a $10\times 10$ grid as inferred from the bias $\rho$ of the corresponding SVM classifiers ($\nu=0.1$). Any two vertices are connected by an edge if $\rho>\rho_c=2.5$.
  The choice of the optimal value of $\rho_c$ is illustrated in the text.
  The phase diagram which has been found based on the peak susceptibilities (cf. Fig.~\ref{fig:D2h_PD}) is shown in red for reference. The phase boundaries are intersected by only few graph edges and this happens mostly where grid points are very close to the phase transition.}
  \label{fig:D2h_graph}
\end{figure}

We revisit the phase diagram for the $D_{2h}$ gauge symmetry of Sec.~\ref{sec:D2h}.
The graph resulting from the above procedure on a $10\times 10$ ($M=100$) grid is shown in Fig.~\ref{fig:D2h_graph}. We used a threshold bias of $\rho_c=2.5$.

Visual inspection of the graph immediately reveals three regions which are densely intraconnected while being sparsely interconnected, corresponding to the three phases of the phase diagram (Fig.~\ref{fig:D2h_PD}). Nevertheless, we point out the utility of spectral graph partitioning methods to identify the phases in an objective fashion. The adjacency matrix $A$ (the $M\times M$ matrix having ones on the off-diagonal elements where vertices are connected by an edge) and degree matrix $D$ (the diagonal matrix where the diagonal elements count the number of edges incident on each vertex) together form the Laplacian matrix $L=D-A$ of the graph. Consequently, the elements of the rows and columns of $L$ sum to zero. When calculating the eigenvalues and eigenvectors of $L$, the smallest eigenvalue is zero and the corresponding eigenvector is proportional to $(1)^{\otimes M}$ for a connected graph. If the graph consists of multiple disconnected components, the eigenvalue $0$ will be degenerate and the eigenvectors reflect the disconnected components.

We use the degeneracy of eigenvalue $0$ to guide our choice of the threshold bias $\rho_c$. We calculate the spectrum of the Laplacian for an assortment of different values for $\rho_c$ and select the largest value for which the eigenvalue $0$ remains nondegenerate. Thus, we tweak $\rho_c$ such that the graph is \emph{just} connected with the understanding that the tentative phase boundary will intersect only few edges. Generically, a higher resolution of the grid will result in more redundant information and the graph will remain connected up to larger values of $\rho_c$. Even larger values of $\rho_c$ would produce disconnected graphs with increasing numbers of components. These could be interpreted as phases directly but the ``desired'' number of phases may not be known \emph{a priori}, so we prefer the following analysis of the (barely) connected graph.

The second smallest eigenvalue of $L$ is referred to as the \emph{algebraic connectivity} and its corresponding eigenvector as the \emph{Fiedler vector}. The elements of the Fiedler vector are clustered around values, positive and negative, and can be used to partition the graph into weakly interconnected components, which in our case correspond to phases.

\begin{figure}
  \centering
  \includegraphics[scale=1.0]{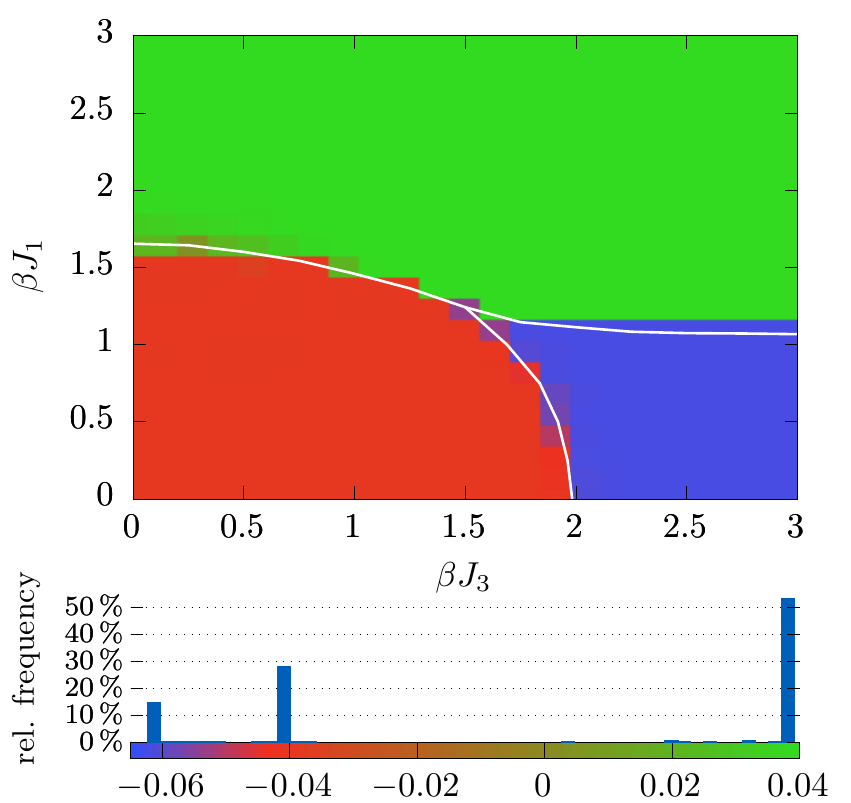}
  \caption{Upper panel: elements of the Fiedler vector corresponding to the graph obtained for a $23\times 23$ grid at a threshold value of $\rho_c=12$. Each element is rendered as a pixel at the location of its corresponding grid point in parameter space and color coded according to its value. The correct phase diagram (cf. Fig.~\ref{fig:D2h_PD}) is shown in white for reference.
  The SVM result shows ambiguity when very close to phase boundaries, however, this tentative phase diagram suffices to guide subsequent trainings to extract order parameters.
  Lower panel: a histogram of the elements' values is shown next to the color scale used to encode them in the upper panel.}
  \label{fig:D2h_phases}
\end{figure}

The elements of the Fiedler vector corresponding to the graph obtained by training $139\,656$ classifiers on a $23\times 23$ grid with threshold bias $\rho_c=12$ is shown in Fig.~\ref{fig:D2h_phases}. At that grid resolution, the graph itself becomes very hard to visualize and is therefore omitted. We histogrammized the elements of the Fiedler vector and identified three distinctive peaks. The color scale has been chosen accordingly, such that the primary colors coincide with those peak values (which are themselves arbitrary). When rendering the Fiedler vector elements in the parameter space at the position of their respective grid points and color-coding them according to that color scale, we reproduce the phase diagram perfectly within the resolution dictated by the grid.

To conclude this section, we remark that in the above cases, only about $10^3$ samples per grid point were used. This is sufficient to obtain a reasonably indicative value for $\rho$. Thus, the total number of samples $N_s$ is comparable to what we typically used in binary of multiclassification. The complexity of the SVM optimization, $\mathcal{O}((N_s/M)^{2.2}M^2)$, is close to that of an equivalent binary classification problem $\mathcal{O}(N_s^{2.2})$. One is also free to choose the grid to explore a more extensive parameter regime or to have higher resolution around features of interest such as the tricritical point. Once the topology of the phase diagram has been extracted, the same samples may be relabeled according to their phase to obtain clean order parameters without the need for resampling.

Lastly, several extensions to this scheme come to mind. Should the computational effort of performing all $M(M-1)/2$ individual binary classification problems turn out to become infeasible, one can limit oneself to binary classification among $n$-th nearest neighbors, thereby exploiting locality of the grid, and reducing the number of binary classification problems from $\mathcal{O}(M^2)$ to $\mathcal{O}(\mathrm{C}_n^2M)$ where $\mathrm{C}_n$ is the number of $n$-th nearest neighbors, $\mathrm{C}_1$ being the coordination number of the grid. (We suggest to use a grid with a high coordination in that case, e.g., a triangular grid which would result in hexagonal pixels.)
Making use of this locality, one will not only gain in computational efficiency, but can also successively apply the multi-classification to parameter regimes of interest, thus effectively explore a large parameter regime of the phase diagram.
Further, it is also possible to combine the graphs obtained by training SVMs with different kernels to capture orders of different ranks, such as encountered for the $D_{3h}$ gauge group in Sec.~\ref{sec:D3h}. Only those edges of the final graph persist which are edges of all constituent graphs. Finally, if the histogram of the Fiedler vector elements exhibits an ambiguous clustering, the alternative approach of \emph{hierarchical clustering} may be taken where the graph is partitioned into two components based on the sign of the Fiedler vector elements and this procedure is iteratively repeated for the Laplacian matrices of the component graphs.

\section{Summary and outlook}\label{sec:conclusion}
In this paper, we have explored multiclassifications using SVMs and extended a kernel method we introduced in a previous work for detecting hidden multipolar orders, to handle multiple phases and extract their order parameters.
We discussed its application in spin and orbital systems and showed in detail the procedure for extracting the analytical order parameter from machine results. We demonstrated this by examining a generalized quadrupolar order and an octupolar order.
The latter is also found in classical kagome antiferromagnets.
There, it took considerable effort to propose its existence~\cite{Ritchey93} and the optimal order parameter~\cite{Zhitomirsky02, Zhitomirsky08}.
However, both this octupolar order and the coexisting quadrupolar order can be identified by our method at once.
While we simulate these orders here with an effective gauge theory, this is only done in light of computational efficiency and flexibility.
The source of the input data to our method is arbitrary, and the kernel~\eqref{eq:kernel} is suitable to detect general classical multipolar orders.

Moreover, we discussed the importance of interpretability in the machine learning of multiple phases.
To physically understand a phase diagram with multiple phases, the goal is not only to distinguish each phase by a symbolic label.
One would also like to know their characterization and transitions.
In addition, in situations where several orders coexist in a phase, which is common in many-body systems, we need to distinguish and identify them individually, as they represent different physics.
This can only be done if the machine is interpretable.

Furthermore, we exploited the internal bias parameter of SVM, and have shown its applications in detecting phase transitions.
Although SVM is technically a supervised learning scheme and demands labeled input data, with this bias parameter, the labeling can be arbitrary and does not need to be faithful to the physical reality.
SVM can differentiate between the case of a phase transition between two sets of state configurations and the case where they were sampled from the same phase.
This can be used as an efficient way to explore unknown phase diagrams.
As we have shown in Sec.~\ref{sec:mapping_PD}, without using such physical information as the order parameter and heat capacity, simply by monitoring the behavior of this bias parameter, SVM can learn a phase diagram with decent quality.
The ambiguity near the phase boundary is acceptable. An approximate phase diagram is sufficient to guide subsequent trainings used for extracting order parameters.
A precise phase diagram can then be computed afterwards.

Our previous work \cite{Greitemann19} and the current paper provide an alternative framework for detecting unconventional orders in frustrated spin and orbital systems.
It may be used as an efficient way for analyzing numerical data and extracting important features from systems of interest.
The current form of the algorithm in principle can be incorporated into any (semi-)classical method where the spin or orbital degrees of freedom can be represented as $\mathrm O(3)$ vectors.
In future work, we plan to extend it to analyze wave functions and density matrices.
However, we note that many frustrated systems already show nontrivial phenomena at the classical level, which also provide useful insight to their quantum counterparts.
Examples can be found in kagome and triangular antiferromagnets, and in volborthite, pyrochlore, and garnet materials.
In particular for three-dimensional systems, owing to the lack of efficient quantum algorithms, their study often largely relies on (semi-)classical methods.

Finally, we note that, as laid out in Sec.~\ref{sec:kernel}, the construction of our kernel is based on properties of local orders and orientational orders, while it is not designed to detect topological quantities.
To the best of our knowledge, machine-learning topological quantities, such as topological defects and topological orders, remains an open problem and does not seem possible yet without feature engineering of the data or ``teaching'' the machine in particular ways.

The source codes and raw data supporting the findings of this study have been made openly available \cite{Jonas}.

\begin{acknowledgements}
This work is supported by FP7/ERC Consolidator Grant No. 771891, the Nanosystems Initiative Munich (NIM), and the Deutsche Forschungsgemeinschaft (DFG, German Research Foundation) under Germany's Excellence Strategy--EXC-2111--390814868.
Our simulations make use of the $\nu$-SVM formulation \cite{Scholkopf00}, the
LIBSVM library \cite{Chang01, Chang11}, and the ALPS\-Core library~
\cite{Gaenko17}.
\end{acknowledgements}

\bibliographystyle{apsrev4-1}
\bibliography{svm2}

\end{document}

%% file: fig_Dinfh_C_0.tex
\begin{tikzpicture}[gnuplot]
\path (0.000,0.000) rectangle (1.000,1.000);
\gpcolor{color=gp lt color border}
\gpsetlinetype{gp lt border}
\gpsetdashtype{gp dt solid}
\gpsetlinewidth{1.00}
\draw[gp path] (0.000,0.999)--(0.000,0.000)--(0.999,0.000)--(0.999,0.999)--cycle;
\gpfill{rgb color={1.000,1.000,1.000}} (0.000,0.888)--(0.111,0.888)--(0.111,0.999)--(0.000,0.999)--cycle;
\gpfill{rgb color={1.000,1.000,1.000}} (0.111,0.888)--(0.222,0.888)--(0.222,0.999)--(0.111,0.999)--cycle;
\gpfill{rgb color={1.000,1.000,1.000}} (0.222,0.888)--(0.333,0.888)--(0.333,0.999)--(0.222,0.999)--cycle;
\gpfill{rgb color={1.000,1.000,1.000}} (0.333,0.888)--(0.444,0.888)--(0.444,0.999)--(0.333,0.999)--cycle;
\gpfill{rgb color={1.000,1.000,1.000}} (0.444,0.888)--(0.555,0.888)--(0.555,0.999)--(0.444,0.999)--cycle;
\gpfill{rgb color={1.000,1.000,1.000}} (0.555,0.888)--(0.666,0.888)--(0.666,0.999)--(0.555,0.999)--cycle;
\gpfill{rgb color={1.000,1.000,1.000}} (0.666,0.888)--(0.777,0.888)--(0.777,0.999)--(0.666,0.999)--cycle;
\gpfill{rgb color={1.000,1.000,1.000}} (0.777,0.888)--(0.888,0.888)--(0.888,0.999)--(0.777,0.999)--cycle;
\gpfill{rgb color={0.000,0.000,0.000}} (0.888,0.888)--(0.999,0.888)--(0.999,0.999)--(0.888,0.999)--cycle;
\gpfill{rgb color={1.000,1.000,1.000}} (0.000,0.777)--(0.111,0.777)--(0.111,0.888)--(0.000,0.888)--cycle;
\gpfill{rgb color={1.000,1.000,1.000}} (0.111,0.777)--(0.222,0.777)--(0.222,0.888)--(0.111,0.888)--cycle;
\gpfill{rgb color={1.000,1.000,1.000}} (0.222,0.777)--(0.333,0.777)--(0.333,0.888)--(0.222,0.888)--cycle;
\gpfill{rgb color={1.000,1.000,1.000}} (0.333,0.777)--(0.444,0.777)--(0.444,0.888)--(0.333,0.888)--cycle;
\gpfill{rgb color={1.000,1.000,1.000}} (0.444,0.777)--(0.555,0.777)--(0.555,0.888)--(0.444,0.888)--cycle;
\gpfill{rgb color={1.000,1.000,1.000}} (0.555,0.777)--(0.666,0.777)--(0.666,0.888)--(0.555,0.888)--cycle;
\gpfill{rgb color={1.000,1.000,1.000}} (0.666,0.777)--(0.777,0.777)--(0.777,0.888)--(0.666,0.888)--cycle;
\gpfill{rgb color={0.000,0.000,0.000}} (0.777,0.777)--(0.888,0.777)--(0.888,0.888)--(0.777,0.888)--cycle;
\gpfill{rgb color={1.000,1.000,1.000}} (0.888,0.777)--(0.999,0.777)--(0.999,0.888)--(0.888,0.888)--cycle;
\gpfill{rgb color={1.000,1.000,1.000}} (0.000,0.666)--(0.111,0.666)--(0.111,0.777)--(0.000,0.777)--cycle;
\gpfill{rgb color={1.000,1.000,1.000}} (0.111,0.666)--(0.222,0.666)--(0.222,0.777)--(0.111,0.777)--cycle;
\gpfill{rgb color={1.000,1.000,1.000}} (0.222,0.666)--(0.333,0.666)--(0.333,0.777)--(0.222,0.777)--cycle;
\gpfill{rgb color={1.000,1.000,1.000}} (0.333,0.666)--(0.444,0.666)--(0.444,0.777)--(0.333,0.777)--cycle;
\gpfill{rgb color={1.000,1.000,1.000}} (0.444,0.666)--(0.555,0.666)--(0.555,0.777)--(0.444,0.777)--cycle;
\gpfill{rgb color={1.000,1.000,1.000}} (0.555,0.666)--(0.666,0.666)--(0.666,0.777)--(0.555,0.777)--cycle;
\gpfill{rgb color={0.000,0.000,0.000}} (0.666,0.666)--(0.777,0.666)--(0.777,0.777)--(0.666,0.777)--cycle;
\gpfill{rgb color={1.000,1.000,1.000}} (0.777,0.666)--(0.888,0.666)--(0.888,0.777)--(0.777,0.777)--cycle;
\gpfill{rgb color={1.000,1.000,1.000}} (0.888,0.666)--(0.999,0.666)--(0.999,0.777)--(0.888,0.777)--cycle;
\gpfill{rgb color={1.000,1.000,1.000}} (0.000,0.555)--(0.111,0.555)--(0.111,0.666)--(0.000,0.666)--cycle;
\gpfill{rgb color={1.000,1.000,1.000}} (0.111,0.555)--(0.222,0.555)--(0.222,0.666)--(0.111,0.666)--cycle;
\gpfill{rgb color={1.000,1.000,1.000}} (0.222,0.555)--(0.333,0.555)--(0.333,0.666)--(0.222,0.666)--cycle;
\gpfill{rgb color={1.000,1.000,1.000}} (0.333,0.555)--(0.444,0.555)--(0.444,0.666)--(0.333,0.666)--cycle;
\gpfill{rgb color={1.000,1.000,1.000}} (0.444,0.555)--(0.555,0.555)--(0.555,0.666)--(0.444,0.666)--cycle;
\gpfill{rgb color={0.000,0.000,0.000}} (0.555,0.555)--(0.666,0.555)--(0.666,0.666)--(0.555,0.666)--cycle;
\gpfill{rgb color={1.000,1.000,1.000}} (0.666,0.555)--(0.777,0.555)--(0.777,0.666)--(0.666,0.666)--cycle;
\gpfill{rgb color={1.000,1.000,1.000}} (0.777,0.555)--(0.888,0.555)--(0.888,0.666)--(0.777,0.666)--cycle;
\gpfill{rgb color={1.000,1.000,1.000}} (0.888,0.555)--(0.999,0.555)--(0.999,0.666)--(0.888,0.666)--cycle;
\gpfill{rgb color={1.000,1.000,1.000}} (0.000,0.444)--(0.111,0.444)--(0.111,0.555)--(0.000,0.555)--cycle;
\gpfill{rgb color={1.000,1.000,1.000}} (0.111,0.444)--(0.222,0.444)--(0.222,0.555)--(0.111,0.555)--cycle;
\gpfill{rgb color={1.000,1.000,1.000}} (0.222,0.444)--(0.333,0.444)--(0.333,0.555)--(0.222,0.555)--cycle;
\gpfill{rgb color={1.000,1.000,1.000}} (0.333,0.444)--(0.444,0.444)--(0.444,0.555)--(0.333,0.555)--cycle;
\gpfill{rgb color={0.000,0.000,0.000}} (0.444,0.444)--(0.555,0.444)--(0.555,0.555)--(0.444,0.555)--cycle;
\gpfill{rgb color={1.000,1.000,1.000}} (0.555,0.444)--(0.666,0.444)--(0.666,0.555)--(0.555,0.555)--cycle;
\gpfill{rgb color={1.000,1.000,1.000}} (0.666,0.444)--(0.777,0.444)--(0.777,0.555)--(0.666,0.555)--cycle;
\gpfill{rgb color={1.000,1.000,1.000}} (0.777,0.444)--(0.888,0.444)--(0.888,0.555)--(0.777,0.555)--cycle;
\gpfill{rgb color={1.000,1.000,1.000}} (0.888,0.444)--(0.999,0.444)--(0.999,0.555)--(0.888,0.555)--cycle;
\gpfill{rgb color={1.000,1.000,1.000}} (0.000,0.333)--(0.111,0.333)--(0.111,0.444)--(0.000,0.444)--cycle;
\gpfill{rgb color={1.000,1.000,1.000}} (0.111,0.333)--(0.222,0.333)--(0.222,0.444)--(0.111,0.444)--cycle;
\gpfill{rgb color={1.000,1.000,1.000}} (0.222,0.333)--(0.333,0.333)--(0.333,0.444)--(0.222,0.444)--cycle;
\gpfill{rgb color={0.000,0.000,0.000}} (0.333,0.333)--(0.444,0.333)--(0.444,0.444)--(0.333,0.444)--cycle;
\gpfill{rgb color={1.000,1.000,1.000}} (0.444,0.333)--(0.555,0.333)--(0.555,0.444)--(0.444,0.444)--cycle;
\gpfill{rgb color={1.000,1.000,1.000}} (0.555,0.333)--(0.666,0.333)--(0.666,0.444)--(0.555,0.444)--cycle;
\gpfill{rgb color={1.000,1.000,1.000}} (0.666,0.333)--(0.777,0.333)--(0.777,0.444)--(0.666,0.444)--cycle;
\gpfill{rgb color={1.000,1.000,1.000}} (0.777,0.333)--(0.888,0.333)--(0.888,0.444)--(0.777,0.444)--cycle;
\gpfill{rgb color={1.000,1.000,1.000}} (0.888,0.333)--(0.999,0.333)--(0.999,0.444)--(0.888,0.444)--cycle;
\gpfill{rgb color={1.000,1.000,1.000}} (0.000,0.222)--(0.111,0.222)--(0.111,0.333)--(0.000,0.333)--cycle;
\gpfill{rgb color={1.000,1.000,1.000}} (0.111,0.222)--(0.222,0.222)--(0.222,0.333)--(0.111,0.333)--cycle;
\gpfill{rgb color={0.000,0.000,0.000}} (0.222,0.222)--(0.333,0.222)--(0.333,0.333)--(0.222,0.333)--cycle;
\gpfill{rgb color={1.000,1.000,1.000}} (0.333,0.222)--(0.444,0.222)--(0.444,0.333)--(0.333,0.333)--cycle;
\gpfill{rgb color={1.000,1.000,1.000}} (0.444,0.222)--(0.555,0.222)--(0.555,0.333)--(0.444,0.333)--cycle;
\gpfill{rgb color={1.000,1.000,1.000}} (0.555,0.222)--(0.666,0.222)--(0.666,0.333)--(0.555,0.333)--cycle;
\gpfill{rgb color={1.000,1.000,1.000}} (0.666,0.222)--(0.777,0.222)--(0.777,0.333)--(0.666,0.333)--cycle;
\gpfill{rgb color={1.000,1.000,1.000}} (0.777,0.222)--(0.888,0.222)--(0.888,0.333)--(0.777,0.333)--cycle;
\gpfill{rgb color={1.000,1.000,1.000}} (0.888,0.222)--(0.999,0.222)--(0.999,0.333)--(0.888,0.333)--cycle;
\gpfill{rgb color={1.000,1.000,1.000}} (0.000,0.111)--(0.111,0.111)--(0.111,0.222)--(0.000,0.222)--cycle;
\gpfill{rgb color={0.000,0.000,0.000}} (0.111,0.111)--(0.222,0.111)--(0.222,0.222)--(0.111,0.222)--cycle;
\gpfill{rgb color={1.000,1.000,1.000}} (0.222,0.111)--(0.333,0.111)--(0.333,0.222)--(0.222,0.222)--cycle;
\gpfill{rgb color={1.000,1.000,1.000}} (0.333,0.111)--(0.444,0.111)--(0.444,0.222)--(0.333,0.222)--cycle;
\gpfill{rgb color={1.000,1.000,1.000}} (0.444,0.111)--(0.555,0.111)--(0.555,0.222)--(0.444,0.222)--cycle;
\gpfill{rgb color={1.000,1.000,1.000}} (0.555,0.111)--(0.666,0.111)--(0.666,0.222)--(0.555,0.222)--cycle;
\gpfill{rgb color={1.000,1.000,1.000}} (0.666,0.111)--(0.777,0.111)--(0.777,0.222)--(0.666,0.222)--cycle;
\gpfill{rgb color={1.000,1.000,1.000}} (0.777,0.111)--(0.888,0.111)--(0.888,0.222)--(0.777,0.222)--cycle;
\gpfill{rgb color={1.000,1.000,1.000}} (0.888,0.111)--(0.999,0.111)--(0.999,0.222)--(0.888,0.222)--cycle;
\gpfill{rgb color={0.000,0.000,0.000}} (0.000,0.000)--(0.111,0.000)--(0.111,0.111)--(0.000,0.111)--cycle;
\gpfill{rgb color={1.000,1.000,1.000}} (0.111,0.000)--(0.222,0.000)--(0.222,0.111)--(0.111,0.111)--cycle;
\gpfill{rgb color={1.000,1.000,1.000}} (0.222,0.000)--(0.333,0.000)--(0.333,0.111)--(0.222,0.111)--cycle;
\gpfill{rgb color={1.000,1.000,1.000}} (0.333,0.000)--(0.444,0.000)--(0.444,0.111)--(0.333,0.111)--cycle;
\gpfill{rgb color={1.000,1.000,1.000}} (0.444,0.000)--(0.555,0.000)--(0.555,0.111)--(0.444,0.111)--cycle;
\gpfill{rgb color={1.000,1.000,1.000}} (0.555,0.000)--(0.666,0.000)--(0.666,0.111)--(0.555,0.111)--cycle;
\gpfill{rgb color={1.000,1.000,1.000}} (0.666,0.000)--(0.777,0.000)--(0.777,0.111)--(0.666,0.111)--cycle;
\gpfill{rgb color={1.000,1.000,1.000}} (0.777,0.000)--(0.888,0.000)--(0.888,0.111)--(0.777,0.111)--cycle;
\gpfill{rgb color={1.000,1.000,1.000}} (0.888,0.000)--(0.999,0.000)--(0.999,0.111)--(0.888,0.111)--cycle;
\draw[gp path] (0.000,0.999)--(0.000,0.000)--(0.999,0.000)--(0.999,0.999)--cycle;
\gpdefrectangularnode{gp plot 1}{\pgfpoint{0.000cm}{0.000cm}}{\pgfpoint{0.999cm}{0.999cm}}
\end{tikzpicture}

%% file: fig_Dinfh_C_1.tex
\begin{tikzpicture}[gnuplot]
\path (0.000,0.000) rectangle (1.000,1.000);
\gpcolor{color=gp lt color border}
\gpsetlinetype{gp lt border}
\gpsetdashtype{gp dt solid}
\gpsetlinewidth{1.00}
\draw[gp path] (0.000,0.999)--(0.000,0.000)--(0.999,0.000)--(0.999,0.999)--cycle;
\gpfill{rgb color={1.000,1.000,1.000}} (0.000,0.888)--(0.111,0.888)--(0.111,0.999)--(0.000,0.999)--cycle;
\gpfill{rgb color={1.000,1.000,1.000}} (0.111,0.888)--(0.222,0.888)--(0.222,0.999)--(0.111,0.999)--cycle;
\gpfill{rgb color={1.000,1.000,1.000}} (0.222,0.888)--(0.333,0.888)--(0.333,0.999)--(0.222,0.999)--cycle;
\gpfill{rgb color={1.000,1.000,1.000}} (0.333,0.888)--(0.444,0.888)--(0.444,0.999)--(0.333,0.999)--cycle;
\gpfill{rgb color={1.000,1.000,1.000}} (0.444,0.888)--(0.555,0.888)--(0.555,0.999)--(0.444,0.999)--cycle;
\gpfill{rgb color={1.000,1.000,1.000}} (0.555,0.888)--(0.666,0.888)--(0.666,0.999)--(0.555,0.999)--cycle;
\gpfill{rgb color={1.000,1.000,1.000}} (0.666,0.888)--(0.777,0.888)--(0.777,0.999)--(0.666,0.999)--cycle;
\gpfill{rgb color={1.000,1.000,1.000}} (0.777,0.888)--(0.888,0.888)--(0.888,0.999)--(0.777,0.999)--cycle;
\gpfill{rgb color={0.000,0.000,0.000}} (0.888,0.888)--(0.999,0.888)--(0.999,0.999)--(0.888,0.999)--cycle;
\gpfill{rgb color={1.000,1.000,1.000}} (0.000,0.777)--(0.111,0.777)--(0.111,0.888)--(0.000,0.888)--cycle;
\gpfill{rgb color={1.000,1.000,1.000}} (0.111,0.777)--(0.222,0.777)--(0.222,0.888)--(0.111,0.888)--cycle;
\gpfill{rgb color={1.000,1.000,1.000}} (0.222,0.777)--(0.333,0.777)--(0.333,0.888)--(0.222,0.888)--cycle;
\gpfill{rgb color={1.000,1.000,1.000}} (0.333,0.777)--(0.444,0.777)--(0.444,0.888)--(0.333,0.888)--cycle;
\gpfill{rgb color={1.000,1.000,1.000}} (0.444,0.777)--(0.555,0.777)--(0.555,0.888)--(0.444,0.888)--cycle;
\gpfill{rgb color={0.000,0.000,0.000}} (0.555,0.777)--(0.666,0.777)--(0.666,0.888)--(0.555,0.888)--cycle;
\gpfill{rgb color={1.000,1.000,1.000}} (0.666,0.777)--(0.777,0.777)--(0.777,0.888)--(0.666,0.888)--cycle;
\gpfill{rgb color={1.000,1.000,1.000}} (0.777,0.777)--(0.888,0.777)--(0.888,0.888)--(0.777,0.888)--cycle;
\gpfill{rgb color={1.000,1.000,1.000}} (0.888,0.777)--(0.999,0.777)--(0.999,0.888)--(0.888,0.888)--cycle;
\gpfill{rgb color={1.000,1.000,1.000}} (0.000,0.666)--(0.111,0.666)--(0.111,0.777)--(0.000,0.777)--cycle;
\gpfill{rgb color={1.000,1.000,1.000}} (0.111,0.666)--(0.222,0.666)--(0.222,0.777)--(0.111,0.777)--cycle;
\gpfill{rgb color={0.000,0.000,0.000}} (0.222,0.666)--(0.333,0.666)--(0.333,0.777)--(0.222,0.777)--cycle;
\gpfill{rgb color={1.000,1.000,1.000}} (0.333,0.666)--(0.444,0.666)--(0.444,0.777)--(0.333,0.777)--cycle;
\gpfill{rgb color={1.000,1.000,1.000}} (0.444,0.666)--(0.555,0.666)--(0.555,0.777)--(0.444,0.777)--cycle;
\gpfill{rgb color={1.000,1.000,1.000}} (0.555,0.666)--(0.666,0.666)--(0.666,0.777)--(0.555,0.777)--cycle;
\gpfill{rgb color={1.000,1.000,1.000}} (0.666,0.666)--(0.777,0.666)--(0.777,0.777)--(0.666,0.777)--cycle;
\gpfill{rgb color={1.000,1.000,1.000}} (0.777,0.666)--(0.888,0.666)--(0.888,0.777)--(0.777,0.777)--cycle;
\gpfill{rgb color={1.000,1.000,1.000}} (0.888,0.666)--(0.999,0.666)--(0.999,0.777)--(0.888,0.777)--cycle;
\gpfill{rgb color={1.000,1.000,1.000}} (0.000,0.555)--(0.111,0.555)--(0.111,0.666)--(0.000,0.666)--cycle;
\gpfill{rgb color={1.000,1.000,1.000}} (0.111,0.555)--(0.222,0.555)--(0.222,0.666)--(0.111,0.666)--cycle;
\gpfill{rgb color={1.000,1.000,1.000}} (0.222,0.555)--(0.333,0.555)--(0.333,0.666)--(0.222,0.666)--cycle;
\gpfill{rgb color={1.000,1.000,1.000}} (0.333,0.555)--(0.444,0.555)--(0.444,0.666)--(0.333,0.666)--cycle;
\gpfill{rgb color={1.000,1.000,1.000}} (0.444,0.555)--(0.555,0.555)--(0.555,0.666)--(0.444,0.666)--cycle;
\gpfill{rgb color={1.000,1.000,1.000}} (0.555,0.555)--(0.666,0.555)--(0.666,0.666)--(0.555,0.666)--cycle;
\gpfill{rgb color={1.000,1.000,1.000}} (0.666,0.555)--(0.777,0.555)--(0.777,0.666)--(0.666,0.666)--cycle;
\gpfill{rgb color={0.000,0.000,0.000}} (0.777,0.555)--(0.888,0.555)--(0.888,0.666)--(0.777,0.666)--cycle;
\gpfill{rgb color={1.000,1.000,1.000}} (0.888,0.555)--(0.999,0.555)--(0.999,0.666)--(0.888,0.666)--cycle;
\gpfill{rgb color={1.000,1.000,1.000}} (0.000,0.444)--(0.111,0.444)--(0.111,0.555)--(0.000,0.555)--cycle;
\gpfill{rgb color={1.000,1.000,1.000}} (0.111,0.444)--(0.222,0.444)--(0.222,0.555)--(0.111,0.555)--cycle;
\gpfill{rgb color={1.000,1.000,1.000}} (0.222,0.444)--(0.333,0.444)--(0.333,0.555)--(0.222,0.555)--cycle;
\gpfill{rgb color={1.000,1.000,1.000}} (0.333,0.444)--(0.444,0.444)--(0.444,0.555)--(0.333,0.555)--cycle;
\gpfill{rgb color={0.000,0.000,0.000}} (0.444,0.444)--(0.555,0.444)--(0.555,0.555)--(0.444,0.555)--cycle;
\gpfill{rgb color={1.000,1.000,1.000}} (0.555,0.444)--(0.666,0.444)--(0.666,0.555)--(0.555,0.555)--cycle;
\gpfill{rgb color={1.000,1.000,1.000}} (0.666,0.444)--(0.777,0.444)--(0.777,0.555)--(0.666,0.555)--cycle;
\gpfill{rgb color={1.000,1.000,1.000}} (0.777,0.444)--(0.888,0.444)--(0.888,0.555)--(0.777,0.555)--cycle;
\gpfill{rgb color={1.000,1.000,1.000}} (0.888,0.444)--(0.999,0.444)--(0.999,0.555)--(0.888,0.555)--cycle;
\gpfill{rgb color={1.000,1.000,1.000}} (0.000,0.333)--(0.111,0.333)--(0.111,0.444)--(0.000,0.444)--cycle;
\gpfill{rgb color={0.000,0.000,0.000}} (0.111,0.333)--(0.222,0.333)--(0.222,0.444)--(0.111,0.444)--cycle;
\gpfill{rgb color={1.000,1.000,1.000}} (0.222,0.333)--(0.333,0.333)--(0.333,0.444)--(0.222,0.444)--cycle;
\gpfill{rgb color={1.000,1.000,1.000}} (0.333,0.333)--(0.444,0.333)--(0.444,0.444)--(0.333,0.444)--cycle;
\gpfill{rgb color={1.000,1.000,1.000}} (0.444,0.333)--(0.555,0.333)--(0.555,0.444)--(0.444,0.444)--cycle;
\gpfill{rgb color={1.000,1.000,1.000}} (0.555,0.333)--(0.666,0.333)--(0.666,0.444)--(0.555,0.444)--cycle;
\gpfill{rgb color={1.000,1.000,1.000}} (0.666,0.333)--(0.777,0.333)--(0.777,0.444)--(0.666,0.444)--cycle;
\gpfill{rgb color={1.000,1.000,1.000}} (0.777,0.333)--(0.888,0.333)--(0.888,0.444)--(0.777,0.444)--cycle;
\gpfill{rgb color={1.000,1.000,1.000}} (0.888,0.333)--(0.999,0.333)--(0.999,0.444)--(0.888,0.444)--cycle;
\gpfill{rgb color={1.000,1.000,1.000}} (0.000,0.222)--(0.111,0.222)--(0.111,0.333)--(0.000,0.333)--cycle;
\gpfill{rgb color={1.000,1.000,1.000}} (0.111,0.222)--(0.222,0.222)--(0.222,0.333)--(0.111,0.333)--cycle;
\gpfill{rgb color={1.000,1.000,1.000}} (0.222,0.222)--(0.333,0.222)--(0.333,0.333)--(0.222,0.333)--cycle;
\gpfill{rgb color={1.000,1.000,1.000}} (0.333,0.222)--(0.444,0.222)--(0.444,0.333)--(0.333,0.333)--cycle;
\gpfill{rgb color={1.000,1.000,1.000}} (0.444,0.222)--(0.555,0.222)--(0.555,0.333)--(0.444,0.333)--cycle;
\gpfill{rgb color={1.000,1.000,1.000}} (0.555,0.222)--(0.666,0.222)--(0.666,0.333)--(0.555,0.333)--cycle;
\gpfill{rgb color={0.000,0.000,0.000}} (0.666,0.222)--(0.777,0.222)--(0.777,0.333)--(0.666,0.333)--cycle;
\gpfill{rgb color={1.000,1.000,1.000}} (0.777,0.222)--(0.888,0.222)--(0.888,0.333)--(0.777,0.333)--cycle;
\gpfill{rgb color={1.000,1.000,1.000}} (0.888,0.222)--(0.999,0.222)--(0.999,0.333)--(0.888,0.333)--cycle;
\gpfill{rgb color={1.000,1.000,1.000}} (0.000,0.111)--(0.111,0.111)--(0.111,0.222)--(0.000,0.222)--cycle;
\gpfill{rgb color={1.000,1.000,1.000}} (0.111,0.111)--(0.222,0.111)--(0.222,0.222)--(0.111,0.222)--cycle;
\gpfill{rgb color={1.000,1.000,1.000}} (0.222,0.111)--(0.333,0.111)--(0.333,0.222)--(0.222,0.222)--cycle;
\gpfill{rgb color={0.000,0.000,0.000}} (0.333,0.111)--(0.444,0.111)--(0.444,0.222)--(0.333,0.222)--cycle;
\gpfill{rgb color={1.000,1.000,1.000}} (0.444,0.111)--(0.555,0.111)--(0.555,0.222)--(0.444,0.222)--cycle;
\gpfill{rgb color={1.000,1.000,1.000}} (0.555,0.111)--(0.666,0.111)--(0.666,0.222)--(0.555,0.222)--cycle;
\gpfill{rgb color={1.000,1.000,1.000}} (0.666,0.111)--(0.777,0.111)--(0.777,0.222)--(0.666,0.222)--cycle;
\gpfill{rgb color={1.000,1.000,1.000}} (0.777,0.111)--(0.888,0.111)--(0.888,0.222)--(0.777,0.222)--cycle;
\gpfill{rgb color={1.000,1.000,1.000}} (0.888,0.111)--(0.999,0.111)--(0.999,0.222)--(0.888,0.222)--cycle;
\gpfill{rgb color={0.000,0.000,0.000}} (0.000,0.000)--(0.111,0.000)--(0.111,0.111)--(0.000,0.111)--cycle;
\gpfill{rgb color={1.000,1.000,1.000}} (0.111,0.000)--(0.222,0.000)--(0.222,0.111)--(0.111,0.111)--cycle;
\gpfill{rgb color={1.000,1.000,1.000}} (0.222,0.000)--(0.333,0.000)--(0.333,0.111)--(0.222,0.111)--cycle;
\gpfill{rgb color={1.000,1.000,1.000}} (0.333,0.000)--(0.444,0.000)--(0.444,0.111)--(0.333,0.111)--cycle;
\gpfill{rgb color={1.000,1.000,1.000}} (0.444,0.000)--(0.555,0.000)--(0.555,0.111)--(0.444,0.111)--cycle;
\gpfill{rgb color={1.000,1.000,1.000}} (0.555,0.000)--(0.666,0.000)--(0.666,0.111)--(0.555,0.111)--cycle;
\gpfill{rgb color={1.000,1.000,1.000}} (0.666,0.000)--(0.777,0.000)--(0.777,0.111)--(0.666,0.111)--cycle;
\gpfill{rgb color={1.000,1.000,1.000}} (0.777,0.000)--(0.888,0.000)--(0.888,0.111)--(0.777,0.111)--cycle;
\gpfill{rgb color={1.000,1.000,1.000}} (0.888,0.000)--(0.999,0.000)--(0.999,0.111)--(0.888,0.111)--cycle;
\draw[gp path] (0.000,0.999)--(0.000,0.000)--(0.999,0.000)--(0.999,0.999)--cycle;
\gpdefrectangularnode{gp plot 1}{\pgfpoint{0.000cm}{0.000cm}}{\pgfpoint{0.999cm}{0.999cm}}
\end{tikzpicture}

%% file: fig_Dinfh_C_2.tex
\begin{tikzpicture}[gnuplot]
\path (0.000,0.000) rectangle (1.000,1.000);
\gpcolor{color=gp lt color border}
\gpsetlinetype{gp lt border}
\gpsetdashtype{gp dt solid}
\gpsetlinewidth{1.00}
\draw[gp path] (0.000,0.999)--(0.000,0.000)--(0.999,0.000)--(0.999,0.999)--cycle;
\gpfill{rgb color={0.000,0.000,0.000}} (0.000,0.888)--(0.111,0.888)--(0.111,0.999)--(0.000,0.999)--cycle;
\gpfill{rgb color={1.000,1.000,1.000}} (0.111,0.888)--(0.222,0.888)--(0.222,0.999)--(0.111,0.999)--cycle;
\gpfill{rgb color={1.000,1.000,1.000}} (0.222,0.888)--(0.333,0.888)--(0.333,0.999)--(0.222,0.999)--cycle;
\gpfill{rgb color={1.000,1.000,1.000}} (0.333,0.888)--(0.444,0.888)--(0.444,0.999)--(0.333,0.999)--cycle;
\gpfill{rgb color={0.000,0.000,0.000}} (0.444,0.888)--(0.555,0.888)--(0.555,0.999)--(0.444,0.999)--cycle;
\gpfill{rgb color={1.000,1.000,1.000}} (0.555,0.888)--(0.666,0.888)--(0.666,0.999)--(0.555,0.999)--cycle;
\gpfill{rgb color={1.000,1.000,1.000}} (0.666,0.888)--(0.777,0.888)--(0.777,0.999)--(0.666,0.999)--cycle;
\gpfill{rgb color={1.000,1.000,1.000}} (0.777,0.888)--(0.888,0.888)--(0.888,0.999)--(0.777,0.999)--cycle;
\gpfill{rgb color={0.000,0.000,0.000}} (0.888,0.888)--(0.999,0.888)--(0.999,0.999)--(0.888,0.999)--cycle;
\gpfill{rgb color={1.000,1.000,1.000}} (0.000,0.777)--(0.111,0.777)--(0.111,0.888)--(0.000,0.888)--cycle;
\gpfill{rgb color={1.000,1.000,1.000}} (0.111,0.777)--(0.222,0.777)--(0.222,0.888)--(0.111,0.888)--cycle;
\gpfill{rgb color={1.000,1.000,1.000}} (0.222,0.777)--(0.333,0.777)--(0.333,0.888)--(0.222,0.888)--cycle;
\gpfill{rgb color={1.000,1.000,1.000}} (0.333,0.777)--(0.444,0.777)--(0.444,0.888)--(0.333,0.888)--cycle;
\gpfill{rgb color={1.000,1.000,1.000}} (0.444,0.777)--(0.555,0.777)--(0.555,0.888)--(0.444,0.888)--cycle;
\gpfill{rgb color={1.000,1.000,1.000}} (0.555,0.777)--(0.666,0.777)--(0.666,0.888)--(0.555,0.888)--cycle;
\gpfill{rgb color={1.000,1.000,1.000}} (0.666,0.777)--(0.777,0.777)--(0.777,0.888)--(0.666,0.888)--cycle;
\gpfill{rgb color={1.000,1.000,1.000}} (0.777,0.777)--(0.888,0.777)--(0.888,0.888)--(0.777,0.888)--cycle;
\gpfill{rgb color={1.000,1.000,1.000}} (0.888,0.777)--(0.999,0.777)--(0.999,0.888)--(0.888,0.888)--cycle;
\gpfill{rgb color={1.000,1.000,1.000}} (0.000,0.666)--(0.111,0.666)--(0.111,0.777)--(0.000,0.777)--cycle;
\gpfill{rgb color={1.000,1.000,1.000}} (0.111,0.666)--(0.222,0.666)--(0.222,0.777)--(0.111,0.777)--cycle;
\gpfill{rgb color={1.000,1.000,1.000}} (0.222,0.666)--(0.333,0.666)--(0.333,0.777)--(0.222,0.777)--cycle;
\gpfill{rgb color={1.000,1.000,1.000}} (0.333,0.666)--(0.444,0.666)--(0.444,0.777)--(0.333,0.777)--cycle;
\gpfill{rgb color={1.000,1.000,1.000}} (0.444,0.666)--(0.555,0.666)--(0.555,0.777)--(0.444,0.777)--cycle;
\gpfill{rgb color={1.000,1.000,1.000}} (0.555,0.666)--(0.666,0.666)--(0.666,0.777)--(0.555,0.777)--cycle;
\gpfill{rgb color={1.000,1.000,1.000}} (0.666,0.666)--(0.777,0.666)--(0.777,0.777)--(0.666,0.777)--cycle;
\gpfill{rgb color={1.000,1.000,1.000}} (0.777,0.666)--(0.888,0.666)--(0.888,0.777)--(0.777,0.777)--cycle;
\gpfill{rgb color={1.000,1.000,1.000}} (0.888,0.666)--(0.999,0.666)--(0.999,0.777)--(0.888,0.777)--cycle;
\gpfill{rgb color={1.000,1.000,1.000}} (0.000,0.555)--(0.111,0.555)--(0.111,0.666)--(0.000,0.666)--cycle;
\gpfill{rgb color={1.000,1.000,1.000}} (0.111,0.555)--(0.222,0.555)--(0.222,0.666)--(0.111,0.666)--cycle;
\gpfill{rgb color={1.000,1.000,1.000}} (0.222,0.555)--(0.333,0.555)--(0.333,0.666)--(0.222,0.666)--cycle;
\gpfill{rgb color={1.000,1.000,1.000}} (0.333,0.555)--(0.444,0.555)--(0.444,0.666)--(0.333,0.666)--cycle;
\gpfill{rgb color={1.000,1.000,1.000}} (0.444,0.555)--(0.555,0.555)--(0.555,0.666)--(0.444,0.666)--cycle;
\gpfill{rgb color={1.000,1.000,1.000}} (0.555,0.555)--(0.666,0.555)--(0.666,0.666)--(0.555,0.666)--cycle;
\gpfill{rgb color={1.000,1.000,1.000}} (0.666,0.555)--(0.777,0.555)--(0.777,0.666)--(0.666,0.666)--cycle;
\gpfill{rgb color={1.000,1.000,1.000}} (0.777,0.555)--(0.888,0.555)--(0.888,0.666)--(0.777,0.666)--cycle;
\gpfill{rgb color={1.000,1.000,1.000}} (0.888,0.555)--(0.999,0.555)--(0.999,0.666)--(0.888,0.666)--cycle;
\gpfill{rgb color={0.000,0.000,0.000}} (0.000,0.444)--(0.111,0.444)--(0.111,0.555)--(0.000,0.555)--cycle;
\gpfill{rgb color={1.000,1.000,1.000}} (0.111,0.444)--(0.222,0.444)--(0.222,0.555)--(0.111,0.555)--cycle;
\gpfill{rgb color={1.000,1.000,1.000}} (0.222,0.444)--(0.333,0.444)--(0.333,0.555)--(0.222,0.555)--cycle;
\gpfill{rgb color={1.000,1.000,1.000}} (0.333,0.444)--(0.444,0.444)--(0.444,0.555)--(0.333,0.555)--cycle;
\gpfill{rgb color={0.000,0.000,0.000}} (0.444,0.444)--(0.555,0.444)--(0.555,0.555)--(0.444,0.555)--cycle;
\gpfill{rgb color={1.000,1.000,1.000}} (0.555,0.444)--(0.666,0.444)--(0.666,0.555)--(0.555,0.555)--cycle;
\gpfill{rgb color={1.000,1.000,1.000}} (0.666,0.444)--(0.777,0.444)--(0.777,0.555)--(0.666,0.555)--cycle;
\gpfill{rgb color={1.000,1.000,1.000}} (0.777,0.444)--(0.888,0.444)--(0.888,0.555)--(0.777,0.555)--cycle;
\gpfill{rgb color={0.000,0.000,0.000}} (0.888,0.444)--(0.999,0.444)--(0.999,0.555)--(0.888,0.555)--cycle;
\gpfill{rgb color={1.000,1.000,1.000}} (0.000,0.333)--(0.111,0.333)--(0.111,0.444)--(0.000,0.444)--cycle;
\gpfill{rgb color={1.000,1.000,1.000}} (0.111,0.333)--(0.222,0.333)--(0.222,0.444)--(0.111,0.444)--cycle;
\gpfill{rgb color={1.000,1.000,1.000}} (0.222,0.333)--(0.333,0.333)--(0.333,0.444)--(0.222,0.444)--cycle;
\gpfill{rgb color={1.000,1.000,1.000}} (0.333,0.333)--(0.444,0.333)--(0.444,0.444)--(0.333,0.444)--cycle;
\gpfill{rgb color={1.000,1.000,1.000}} (0.444,0.333)--(0.555,0.333)--(0.555,0.444)--(0.444,0.444)--cycle;
\gpfill{rgb color={1.000,1.000,1.000}} (0.555,0.333)--(0.666,0.333)--(0.666,0.444)--(0.555,0.444)--cycle;
\gpfill{rgb color={1.000,1.000,1.000}} (0.666,0.333)--(0.777,0.333)--(0.777,0.444)--(0.666,0.444)--cycle;
\gpfill{rgb color={1.000,1.000,1.000}} (0.777,0.333)--(0.888,0.333)--(0.888,0.444)--(0.777,0.444)--cycle;
\gpfill{rgb color={1.000,1.000,1.000}} (0.888,0.333)--(0.999,0.333)--(0.999,0.444)--(0.888,0.444)--cycle;
\gpfill{rgb color={1.000,1.000,1.000}} (0.000,0.222)--(0.111,0.222)--(0.111,0.333)--(0.000,0.333)--cycle;
\gpfill{rgb color={1.000,1.000,1.000}} (0.111,0.222)--(0.222,0.222)--(0.222,0.333)--(0.111,0.333)--cycle;
\gpfill{rgb color={1.000,1.000,1.000}} (0.222,0.222)--(0.333,0.222)--(0.333,0.333)--(0.222,0.333)--cycle;
\gpfill{rgb color={1.000,1.000,1.000}} (0.333,0.222)--(0.444,0.222)--(0.444,0.333)--(0.333,0.333)--cycle;
\gpfill{rgb color={1.000,1.000,1.000}} (0.444,0.222)--(0.555,0.222)--(0.555,0.333)--(0.444,0.333)--cycle;
\gpfill{rgb color={1.000,1.000,1.000}} (0.555,0.222)--(0.666,0.222)--(0.666,0.333)--(0.555,0.333)--cycle;
\gpfill{rgb color={1.000,1.000,1.000}} (0.666,0.222)--(0.777,0.222)--(0.777,0.333)--(0.666,0.333)--cycle;
\gpfill{rgb color={1.000,1.000,1.000}} (0.777,0.222)--(0.888,0.222)--(0.888,0.333)--(0.777,0.333)--cycle;
\gpfill{rgb color={1.000,1.000,1.000}} (0.888,0.222)--(0.999,0.222)--(0.999,0.333)--(0.888,0.333)--cycle;
\gpfill{rgb color={1.000,1.000,1.000}} (0.000,0.111)--(0.111,0.111)--(0.111,0.222)--(0.000,0.222)--cycle;
\gpfill{rgb color={1.000,1.000,1.000}} (0.111,0.111)--(0.222,0.111)--(0.222,0.222)--(0.111,0.222)--cycle;
\gpfill{rgb color={1.000,1.000,1.000}} (0.222,0.111)--(0.333,0.111)--(0.333,0.222)--(0.222,0.222)--cycle;
\gpfill{rgb color={1.000,1.000,1.000}} (0.333,0.111)--(0.444,0.111)--(0.444,0.222)--(0.333,0.222)--cycle;
\gpfill{rgb color={1.000,1.000,1.000}} (0.444,0.111)--(0.555,0.111)--(0.555,0.222)--(0.444,0.222)--cycle;
\gpfill{rgb color={1.000,1.000,1.000}} (0.555,0.111)--(0.666,0.111)--(0.666,0.222)--(0.555,0.222)--cycle;
\gpfill{rgb color={1.000,1.000,1.000}} (0.666,0.111)--(0.777,0.111)--(0.777,0.222)--(0.666,0.222)--cycle;
\gpfill{rgb color={1.000,1.000,1.000}} (0.777,0.111)--(0.888,0.111)--(0.888,0.222)--(0.777,0.222)--cycle;
\gpfill{rgb color={1.000,1.000,1.000}} (0.888,0.111)--(0.999,0.111)--(0.999,0.222)--(0.888,0.222)--cycle;
\gpfill{rgb color={0.000,0.000,0.000}} (0.000,0.000)--(0.111,0.000)--(0.111,0.111)--(0.000,0.111)--cycle;
\gpfill{rgb color={1.000,1.000,1.000}} (0.111,0.000)--(0.222,0.000)--(0.222,0.111)--(0.111,0.111)--cycle;
\gpfill{rgb color={1.000,1.000,1.000}} (0.222,0.000)--(0.333,0.000)--(0.333,0.111)--(0.222,0.111)--cycle;
\gpfill{rgb color={1.000,1.000,1.000}} (0.333,0.000)--(0.444,0.000)--(0.444,0.111)--(0.333,0.111)--cycle;
\gpfill{rgb color={0.000,0.000,0.000}} (0.444,0.000)--(0.555,0.000)--(0.555,0.111)--(0.444,0.111)--cycle;
\gpfill{rgb color={1.000,1.000,1.000}} (0.555,0.000)--(0.666,0.000)--(0.666,0.111)--(0.555,0.111)--cycle;
\gpfill{rgb color={1.000,1.000,1.000}} (0.666,0.000)--(0.777,0.000)--(0.777,0.111)--(0.666,0.111)--cycle;
\gpfill{rgb color={1.000,1.000,1.000}} (0.777,0.000)--(0.888,0.000)--(0.888,0.111)--(0.777,0.111)--cycle;
\gpfill{rgb color={0.000,0.000,0.000}} (0.888,0.000)--(0.999,0.000)--(0.999,0.111)--(0.888,0.111)--cycle;
\draw[gp path] (0.000,0.999)--(0.000,0.000)--(0.999,0.000)--(0.999,0.999)--cycle;
\gpdefrectangularnode{gp plot 1}{\pgfpoint{0.000cm}{0.000cm}}{\pgfpoint{0.999cm}{0.999cm}}
\end{tikzpicture}